\documentclass[12pt]{iopart}
\usepackage{amsfonts}
\usepackage{amssymb}
\usepackage[usenames, dvipsnames]{color}
\usepackage{graphicx}
\usepackage{leftidx}
\usepackage[breaklinks,colorlinks = true,linkcolor = red,urlcolor=cyan,citecolor=red]{hyperref}
\usepackage{mathrsfs}
\usepackage{ulem}
\usepackage{bm}

\DeclareMathAlphabet\mathbfcal{OMS}{cmsy}{b}{n}
\newcommand{\imineq}[2]{\vcenter{\hbox{\includegraphics[height=#2ex]{#1}}}}

\begin{document}

\title{Signatures of spin-triplet excitations in optical conductivity of valence bond solids}
\date{\today}

\author{Kyusung Hwang$^1$, Subhro Bhattacharjee$^1$\footnote{Present Address: Max-Planck Institute for the Physics of Complex Systems, Nothnitzer Str. 38, 01187 Dresden, Germany.}, Yong Baek Kim$^{1,2}$}

\address{$^1$Department of Physics and Centre for Quantum Materials, University of Toronto, Toronto, Ontario M5S 1A7, Canada, \\
$^2$School of Physics, Korea Institute for Advanced Study, Seoul 130-722, Korea}

\eads{\mailto{khwang@physics.utoronto.ca},
\mailto{subhro@pks.mpg.de},  \mailto{ybkim@physics.utoronto.ca}}

\begin{abstract}
We show that the optical responses below the Mott gap can be used to probe the spin-triplet excitations in valence bond solid (VBS) phases in Mott insulators. The optical conductivity in this regime arises due to the electronic polarization mechanism via virtual electron hopping processes. We apply this mechanism to the Hubbard model with spin-orbit couplings and/or the corresponding spin model with significant Dzyaloshinskii-Moriya (DM) interactions, and compute the optical conductivity of VBS states on both ideal and deformed Kagome lattices. In case of the deformed Kagome lattice, we study the antiferromagnet, Rb$_2$Cu$_3$SnF$_{12}$ with the pinwheel VBS state. In case of the ideal Kagome lattice, we explore the optical conductivity signatures of the spin-triplet excitations for three VBS states with (1) a 12-site unit cell, (2) a 36-site unit cell with six-fold rotation symmetry, and (3) a 36-site unit cell with three-fold rotation symmetry, respectively. We find that increasing the DM interactions generally leads to broad and smooth features in the optical conductivity with interesting experimental consequences. The optical conductivity reflects the features of the spin-triplet excitations that can be measured in future experiments. 
\end{abstract}

\maketitle

\section{Introduction\label{sec:intro}}

In Mott insulators, small charge fluctuations exist due to virtual hopping of  the electrons for any finite hopping amplitude. It is well-known that these virtual hoppings generate the spin-spin exchange interactions in the Heisenberg type spin Hamiltonians that describe the low energy physics of Mott insulators at half-filling. Bulaevskii {\it et al.} \cite{2008_Bulaevskii} have shown that such virtual charge fluctuations may also lead to non-zero electronic polarization resulting in finite charge response even inside the insulating regime.

Such charge polarization leads to, for example, finite optical conductivity much below the single-particle charge gap whose measurement can yield useful information about the low energy physics of Mott insulators. While such measurements may be generically interesting, probing the optical conductivity of Mott insulators that do not exhibit magnetic order at low temperatures-- the quantum paramagnets, can be particularly worthwhile in providing insight into the magnetically disordered ground states. Given that some of these quantum paramagnets and associated phase transitions are presently far from well-understood, understanding the efficacy of such probes as optical conductivity, is worth exploring. Recently, some of these ideas have been studied in context of both U(1) and Z$_2$ spin liquids and related phase transitions in spin-1/2 Kagome lattice antiferromagnets \cite{2013_Potter,2013_Huh}. On the experimental side, interesting power-law behavior was observed in recent optical conductivity measurements \cite{2013_Pilon} on the Herbertsmithite [ZnCu$_3$(OH)$_6$Cl$_2$], a spin-1/2 antiferromagnet on an ideal Kagome lattice, which is widely believed to realize a spin liquid ground state \cite{2007_Mendels,2007_Helton,2008_Zorko,2012_Han}.

An interesting result obtained by Bulaevskii {\it et al.} \cite{2008_Bulaevskii} is that, in a Mott insulator with spin rotation symmetry, to the leading order, the magnitude of the effective electron polarization operator at a site $i$ is given by $|\mathbfcal{P}_i|\sim ({\bf S}_j \cdot {\bf S}_k - {\bf S}_i \cdot {\bf S}_j) + ({\bf S}_j \cdot {\bf S}_k - {\bf S}_i \cdot {\bf S}_k)$, where $i,j,k$ form an elementary triangle of underlying lattice. However, this is nothing but the local operator measuring formation of spin-singlets (dimers) on the triangle. Hence, such a dimer operator can directly couple to an external electric field. It is therefore interesting to ask about the nature of the subgap (below charge gap) optical conductivity in different ground states that show such dimerization. We shall refer to such states as valence bond solids (VBS). 

In this paper, we explore the charge response of various such VBS states on both ideal and deformed Kagome lattices by calculating the subgap optical conductivity. While the calculations, in principle, can be extended to any lattice, we choose to concentrate on the Kagome lattice generally because of  the large number of interesting compounds that have been investigated on both ideal and deformed Kagome lattices \cite{2007_Mendels,2007_Helton,2008_Zorko,2012_Han,2001_Hiroi,2009_Okamoto,2013_Clark,2008_Morita,2010_Matan}. In the deformed Kagome lattice, like in the case of Rb$_2$Cu$_3$SnF$_{12}$ \cite{2008_Morita,2010_Matan,2009_Yang} (see Fig. \ref{fig:deformed_kagome_lattice}) such VBS order has already been found. Also, recent variational Monte Carlo simulations  suggest that VBS order may be stabilized in spin-1/2 Heisenberg antiferromagnets on an ideal Kagome lattice in presence of small further neighbour couplings \cite{2011_Iqbal}.

Many of the above experimentally relevant Mott insulators are described by spin Hamiltonians with non-negligible Dzyaloshinskii-Moriya (DM) interactions \cite{1958_Dzyaloshinskii,1960_Moriya}. Hence, in this paper, we extend the previous results in Ref. \cite{2008_Bulaevskii}, and derive the form of the low energy electronic polarization operator in presence of such DM interactions. Using a bond operator mean-field theory \cite{1990_Sachdev,1994_Gopalan,2009_Yang,2012_Hwang} that captures the low energy gapped spin-triplet excitations (triplons) in a VBS phase, we then calculate the leading order triplon contribution to the optical conductivity. This leading order contribution comes from the two triplon excitations. Hence, the optical response have a finite gap characterized by the minimum of two-triplon excitation energy. We find that incorporation of the DM interactions has a pronounced effect on the optical conductivity (see Fig. \ref{fig:VBS_patterns}, \ref{fig:12_VBS_band_conductivity}, \ref{fig:36_VBS_6_band_conductivity}, \ref{fig:36_VBS_3_band_conductivity}). As the  strength of the DM term is increased, sharp features in the optical conductivity are generally replaced by a smoother and broader structures. This may actually lead to interesting consequences as the broad structure in the optical conductivity may mimic power-law optical responses expected in U(1) spin liquids \cite{2013_Potter,2013_Pilon}, albeit above finite gap, over short range of experimental measurement. Interestingly, similar ``smoothening" of sharper structures due to the DM interactions have also been seen in calculations of the dynamic spin structure factor and electron spin resonance (ESR) spectra of a large number of U(1) and Z$_2$ spin liquids on the ideal Kagome lattice \cite{2013_Dodds}. Near a transition between a VBS and a magnetically ordered state, which is brought about by the condensation of the triplon, the triplon gap closes and hence the optical response  should be observed at very low frequencies. 

As a concrete example of VBS order, we take the deformed Kagome lattice antiferromagnet Rb$_2$Cu$_3$SnF$_{12}$, which has a 12-site unit-cell VBS ground state with a characteristic {\it pinwheel} structure \cite{2008_Morita,2010_Matan,2009_Yang} (see Fig. \ref{fig:deformed_kagome_lattice}). Using  a triplon spectra that matches with the experimental energy scales, we calculate the optical conductivity and study its characteristic features. Our calculation suggests that the lower bound of the optical conductivity response in Rb$_2$Cu$_3$SnF$_{12}$ lies in the THz frequency regime with an intensity of the order of $10^{-6} \sim 10^{-4} ~\Omega^{-1}\textup{cm}^{-1}$. In case of VBS orders on the ideal Kagome lattice, we study three well known VBS states that have been studied in different contexts \cite{2011_Huh,1991_Marston,2003_Nikolic,2007_Singh}: (1) a 12-site pinwheel VBS, (2) a 36-site VBS with six-fold rotation symmetry, and (3) a 36-site VBS with three-fold rotation symmetry.

In addition to the virtual-charge fluctuation mechanism discussed above, additional contribution to the effective electric polarization can occur due to magneto-elastic coupling and also lead to finite subgap optical conductivity \cite{2013_Potter,2013_Huh}. While we do not consider the effect of such magneto-elastic coupling in this work, we shall comment on its effects towards the end of the paper.

The rest of the paper is organized as follows. In Sec. \ref{sec:Optical_response_of_Mott_insulator}, we generalize and extend the work by Bulaevskii {\it et al.} to derive effective electronic polarization operator in the case where the DM interactions are present in low energy spin Hamiltonian.  With this polarization operator, a linear response theory is then developed for the subgap optical conductivity in Mott insulators.  In Sec. \ref{sec:kagome_optical_conductivity}, we introduce the spin model on the Kagome lattice that will be considered throughout the paper and provide a recipe to construct the polarization operator based on symmetries of the model. After a brief review on the bond operator mean-field theory for VBS order in Sec. \ref{sec:bond_op_optical_conductivity}, the polarization and optical conductivity are re-expressed in terms of the bond operators. The calculation of the optical conductivity with the triplon excitations and discussion of the results are presented for the pinwheel VBS state in Rb$_2$Cu$_3$SnF$_{12}$ in Sec. \ref{sec:deformed_kagome_lattice} and for various VBS orders on the ideal Kagome lattice in Sec. \ref{sec:ideal_kagome_lattice}. We summarize our results and discuss possible implications with regards to various experiments in Sec. \ref{sec:last_sec}.


\section{Optical response of Mott insulator: electronic polarization mechanism\label{sec:Optical_response_of_Mott_insulator}}

Bulaevskii {\it et al.} \cite{2008_Bulaevskii} showed that in a Mott insulator, described by the large $U$ limit of the spin-rotation invariant Hubbard model, the virtual charge fluctuations can lead to finite electronic polarization which can then couple to an external electric field giving rise to finite optical response. 

However, many interesting experimental examples of Mott insulators actually break the spin-rotation symmetry due to the presence of atomic spin-orbit coupling. So, it is important to ask for the effect of these symmetry breaking terms on the optical response. Indeed, we find that such terms have characteristic contributions, which we derive by starting from an appropriate Hubbard model in Eq. (\ref{eq:Hubbard_model}) and constructing effective spin Hamiltonian and polarization operators to the leading order of perturbation theory in the large $U$ limit. To this end, we re-derive the results by Bulaevski {\it et al.} \cite{2008_Bulaevskii} with appropriate generalizations.

\subsection{Large $U$ limit: the spin Hamiltonian and electronic polarization operator}

We start with a single band Hubbard model at half-filling:
\begin{equation}
 H = H_h + H_U = \sum_{i,j} c_{i\alpha}^{\dagger} h_{ij,\alpha\beta} c_{j\beta} + U \sum_{i} n_{i\uparrow} n_{i\downarrow},
 \label{eq:Hubbard_model}
\end{equation}
where $c_{i\alpha}^{\dagger}$ ($c_{j\beta}$) is electron creation (annihilation) operator ($i,j$ are site indices and $\alpha,\beta(=\uparrow,\downarrow)$ are spin indices) and $n_{i\alpha}=c_{i\alpha}^{\dagger}c_{i\alpha}$. The hopping parameters $h_{ij,\alpha\beta}=t_{ij} \delta_{\alpha\beta} + i {\bf v}_{ij} \cdot \boldsymbol{\sigma}_{\alpha\beta}$ are in general spin-dependent in presence of spin-orbit coupling. Here $\delta_{\alpha\beta}$ is the Kronecker delta and $\boldsymbol{\sigma}$ are the Pauli matrices. The parameters $t_{ij}$ and ${\bf v}_{ij}$ are real scalars and real pseudo-vectors respectively which are constrained by the hermiticity as well as time-reversal symmetry of the Hamiltonian : $t_{ji}=t_{ij},~{\bf v}_{ji}=-{\bf v}_{ij}$. The parameters can be further constrained by the space group symmetries of the lattice under consideration. At the moment, we do not assume any particular lattice.

Since there are no charge-current carrying bulk states in the  Mott insulator, the optical response is generated due to finite polarization of the system. There are several independent contributions to the polarization which has two chief sources, the electrons and the phonons. Hereafter, we assume a rigid lattice and concentrate on the electronic part of the polarization. We shall briefly comment on the phonon contribution towards the end (Sec. \ref{sec:last_sec}). The electronic contribution of the polarization operator is given by:
\begin{equation}
 {\bf P} = \frac{1}{V} \sum_{i} e \delta n_i {\bf r}_i,
 \label{eq:P}
\end{equation}
where $e(<0)$ is the electron charge, $\delta n_i = n_i-1 ~(n_i=n_{i\uparrow}+n_{i\downarrow})$ is the electron number fluctuation from single occupancy at site $i$ with position ${\bf r}_i$, and $V$ is the volume of the system. The above polarization operator indicates deviation from the charge neutrality caused by virtual fluctuation in the electron number at each site.

The low energy physics of Mott insulator is described by effective spin Hamiltonians that can be obtained through well known strong-coupling expansion in the small $h_{ij}/U$ limit \cite{1988_MacDonald}. As noted before, the same virtual charge fluctuations that give rise to the low energy spin-dynamics also lead to non-zero effective electronic polarization \cite{2008_Bulaevskii}. To be specific, in the strong-coupling expansion, the effective spin Hamiltonian $\mathcal{H}$ and electronic polarization $\mathbfcal{P}$ are constructed in the degenerate ground state manifold of the Hubbard model $H$ with $h_{ij}=0$, with single occupancy of electron at each site;
\begin{equation}
 \mathcal{H} = P_s e^{\mathcal{S}} H e^{-\mathcal{S}} P_s ,
\end{equation}
\begin{equation}
 \mathbfcal{P} = P_s e^{\mathcal{S}} {\bf P} e^{-\mathcal{S}} P_s.
\end{equation}
Here, $P_s$ is the projection into the ground state manifold, and $e^{-\mathcal{S}}$ is the unitary transformation ($\mathcal{S}$ is anti-Hermitian) that transforms states $\{ | \psi \rangle \}$ in the manifold into low energy eigenstates $\{ | \phi \rangle \}$ of $H$, {\it i.e.} $| \phi \rangle = e^{-\mathcal{S}} | \psi \rangle$.
The unitary transformation operator contains virtual hopping effects on the ground state manifold and can be expanded perturbatively in terms of $h_{ij}/U$.
In general, the above applies to any physical operator $O$ and its effective counterpart $\mathcal{O}$ with the relationship $\mathcal{O} = P_s e^{\mathcal{S}} O e^{-\mathcal{S}} P_s$, which is equivalent to $\langle \psi | \mathcal{O} | \psi' \rangle = \langle \phi | O | \phi' \rangle$.
Below, we present the forms of effective spin Hamiltonian and electronic polarization operator in presence of spin-orbit coupling.
\subsubsection{Low energy spin Hamiltonian :}

To the leading (second) order, the well-known low energy effective spin Hamiltonian $\mathcal{H}$ consists of the Heisenberg ($J_{ij}$), Dzyaloshinskii-Moriya (${\bf D}_{ij}$), and anisotropic (${\Gamma_{ij}^{ab}}$) interactions:
\begin{equation}
\mathcal{H} =
\sum_{i,j} 
\left(
J_{ij} {\bf S}_i \cdot {\bf S}_j 
+ 
{\bf D}_{ij} \cdot {\bf S}_i \times {\bf S}_j 
+
S_i^a \Gamma_{ij}^{ab} S_j^b
\right),
\label{eq:H_eff_final_form}
\end{equation}
where ${\bf S}_i=\frac{1}{2}c_{i\alpha}^{\dagger}{\sigma}_{\alpha\beta}c_{i\beta}$, $a,b=x,y,z$. The coupling constants are given by
\begin{equation}
 J_{ij}=\frac{4(t_{ij}^2-{\bf v}_{ij}^2/3)}{U},
 \\
 {\bf D}_{ij} = \frac{8t_{ij}{\bf v}_{ij}}{U},
 \\
 \Gamma_{ij}^{ab}=\frac{8(v_{ij}^a v_{ij}^b - \delta^{ab} {\bf v}_{ij}^2/3)}{U}.
\end{equation}
The higher order corrections are subleading in powers of $h_{ij}/U$ and deep inside the Mott insulator, their effects are negligible.
\subsubsection{Effective electronic polarization operator :\label{sec:P_eff}}
The lowest order nonzero corrections to the electronic polarization operator occur at the {\it third} order of the strong-coupling expansion.
This leading order contributions are present in systems where the hoppings are nonzero over elementary triangles such as in the triangular and Kagome lattices.
The effective electron number fluctuation $\delta \mathcal{N}_i=P_s e^{\mathcal{S}} \delta n_i e^{-\mathcal{S}} P_s$ in the large $U$ limit at a site $i$ on a triangle $\imineq{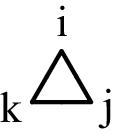}{5}$ has the following form:
\begin{eqnarray}
 \delta \mathcal{N}_i
 &=& - A_{ijk} ( {\bf S}_i \cdot {\bf S}_j - {\bf v}'_{ij,k} \cdot {\bf S}_i \times {\bf S}_j )
 + 2 A_{ijk} ( {\bf S}_j \cdot {\bf S}_k - {\bf v}'_{jk,i} \cdot {\bf S}_j \times {\bf S}_k )\nonumber\\
 && - A_{ijk} ( {\bf S}_k \cdot {\bf S}_i - {\bf v}'_{ki,j} \cdot {\bf S}_k \times {\bf S}_i ),
 \label{eq:n_i_eff_final_form}
\end{eqnarray}
where
\begin{equation}
 A_{ijk}=\frac{8t_{ij} t_{jk} t_{ki}}{U^3},~~~~
 {\bf v}'_{ij,k}=-\frac{{\bf v}_{ij}}{t_{ij}}+\frac{{\bf v}_{jk}}{t_{jk}}+\frac{{\bf v}_{ki}}{t_{ki}},
\end{equation}
and ${\bf v}'_{jk,i}$ and ${\bf v}'_{ki,j}$ are obtained by cyclic permutations of $ijk$ in ${\bf v}'_{ij,k}$.
The above form of $\delta \mathcal{N}_i$ only describes charge fluctuation on a given triangle $ijk$. If the site $i$ belongs to several triangles as in the triangular and Kagome lattices, all possible triangular charge fluctuations should be combined for the correct form of $\delta \mathcal{N}_i$.
Keeping this in mind, we construct the effective polarization operator as
\begin{equation}
 \mathbfcal{P}=\frac{1}{V} \sum_{i} e \delta \mathcal{N}_i {\bf r}_i.
\label{eq:P_eff}
\end{equation}
It is worthwhile to note that the effective spin Hamiltonian and the polarization operators have their first non-trivial contributions at the second and third orders of the strong-coupling expansion, respectively.
\subsection{Linear response theory for optical conductivity}

The above polarization operator can couple to an external electric field linearly: $- V{\mathbfcal{P}} \cdot {\bf E}(t)$, where ${\bf E}(t)$ is a time-dependent, spatially uniform external electric field. Within linear response theory, the electric susceptibility in frequency space is
\begin{eqnarray}
 \chi_{ab}(\omega) =
 - \frac{V}{\hbar}\left[ \sum_{n\ne0} \frac{\langle \psi_0 | \mathcal{P}_{a} | \psi_n \rangle \langle \psi_n | \mathcal{P}_{b} | \psi_0 \rangle}{\omega-\omega_{n}+i0^{+}}- \sum_{n\ne0} \frac{\langle \psi_0 | \mathcal{P}_{b} | \psi_n \rangle \langle \psi_n | \mathcal{P}_{a} | \psi_0 \rangle}{\omega+\omega_{n}+i0^{+}} \right],
 \label{eq:chi}
\end{eqnarray}
where  {$\mathcal{H} | \psi_n \rangle = E_n | \psi_n \rangle ~ (n=0,1,\cdots)$} with $E_0 < E_1 < \cdots$ with $| \psi_0 \rangle$ being the ground state, and $\omega_{n}=(E_n-E_0)/\hbar$. The optical conductivity $\sigma_{ab}(\omega)$ is then given by the standard relation \cite{Marder}
\begin{equation}
 \sigma_{ab}(\omega) = - i \omega \chi_{ab}(\omega), ~~~~~{\it i.e.}~~~~~~~ \textup{Re}[\sigma_{ab}(\omega)] = \omega \textup{Im}[\chi_{ab}(\omega)]
 \label{eq:Re_sigma}
\end{equation}
Some comments regarding the above expression are essential. The above expression is really valid in the frequency regime much below the energy scale associated with single-particle charge excitation in the Hubbard model (Eq. \ref{eq:Hubbard_model}). This latter scale is in the order of $U$. Only in this regime, the optical response is correctly captured by the coupling of the effective polarization, $\mathbfcal{P}$, to the external electric field. Hence, our calculation is valid in the regime $\hbar \omega \ll U$.

\section{Optical conductivity for Valence bond solids in Kagome lattice\label{sec:kagome_optical_conductivity}}

We shall now apply the above linear response theory to calculate optical conductivity in various VBS states on the Kagome lattice. In all the systems, we shall concentrate on cases where the spin-dependent hopping part is small compared to the spin-independent hopping part ($|{\bf v}_{ij}|/t_{ij} \ll 1$) so that the $\Gamma^{ab}_{ij}$ terms can be neglected. Further, we shall restrict our attention to the cases where the DM vectors point perpendicular to the Kagome plane, {\it i.e.} ${\bf D}_{ij}=D_{ij}\hat{z}$ (we set the Kagome lattice to lie in the $xy$ plane). The resultant Hamiltonian is then given by:
\begin{equation}
\mathcal{H} 
=
\sum_{i,j} 
\left(
J_{ij}
{\bf S}_i \cdot {\bf S}_j 
+ 
D_{ij} \hat{z} \cdot {\bf S}_i \times {\bf S}_j 
\right).
\label{eq:H_eff_without_Gamma}
\end{equation}
The Hamiltonian has following symmetries: (1) time-reversal, (2) lattice-translation, (3) SO(2) spin-rotation along the $z$ axis, and, (4) six-fold lattice-rotation $C_6$ and inversion $I$ with respect to the centers of certain hexagons (see later  for more details). Throughout our calculations, we shall make use of these and other symmetries which arise for special parameter values.

The VBS ground states do not break the time-reversal symmetry of the Hamiltonian. This immediately implies that the off-diagonal components ($a \ne b$) of the optical conductivity are identically zero. Further, all the VBS states that we consider either have a $C_6$ or $C_3$ rotation symmetry. In absence of off-diagonal terms, such $C_{3,6}$ rotation symmetry implies that the optical response is isotropic. Therefore, from  Eqs. (\ref{eq:chi}) and (\ref{eq:Re_sigma}) with $\omega>0$ we have
\begin{equation}
\textup{Re}[\sigma_{xx}(\omega)] = \textup{Re}[\sigma_{yy}(\omega)] \equiv \sigma (\omega).
\end{equation}
The isotropic optical conductivity $\sigma (\omega)$ is given by following formula:
\begin{equation}
 \sigma(\omega)=
 \frac{\pi V\omega}{\hbar} \sum_{n\ne0} 
 | \langle \psi_n | \mathcal{P}_{a} | \psi_0 \rangle |^2 \delta (\omega - \omega_{n}).
 ~~~
 \label{eq:Re_sigma_final}
\end{equation}

Since the optical conductivity calculation will be done with exchange couplings in the spin Hamiltonian, by using the relationships $J_{ij} \simeq 4t_{ij}^2/U$ and ${\bf v}_{ij}/t_{ij} \simeq {\bf D}_{ij}/2J_{ij}=D_{ij}\hat{z}/2J_{ij}$ we rewrite the coefficients in the effective number operator (\ref{eq:n_i_eff_final_form}) in terms of $J_{ij},D_{ij}$:
\begin{equation}
 A_{ijk} \simeq \sqrt{\frac{J_{ij}J_{jk}J_{ki}}{U^3}},
 ~~~~~
{\bf v}'_{ij,k} \simeq \frac{d_{ij,k}}{2}\hat{z},
\end{equation}
where
\begin{equation}
d_{ij,k} = -\frac{D_{ij}}{J_{ij}}+\frac{D_{jk}}{J_{jk}}+\frac{D_{ki}}{J_{ki}}.
\end{equation}
Then, with the coefficients we obtain following simple expression for the electron number fluctuation operator:
\begin{equation}
 \delta \mathcal{N}_i
 =
 A_{ijk} ( - \mathcal{I}_{ij} + 2 \mathcal{I}_{jk} - \mathcal{I}_{ki} ),
 \label{eq:I_ij}
\end{equation}
with
\begin{equation}
 \mathcal{I}_{ij}={\bf S}_i \cdot {\bf S}_j - \frac{d_{ij,k}}{2} \hat{z} \cdot {\bf S}_i \times {\bf S}_j.
 \label{eq:I_ij}
\end{equation}
Here, the site $k$ in $d_{ij,k}$ is determined uniquely for a given link $ij$ since the bond lies in only one triangle on the Kagome lattice.
It must be noted that the above expression of $\delta \mathcal{N}_i$ is only valid for one triangle $ijk$. For the accurate expression, contributions from all triangles having the site $i$ must be summed up in $\delta \mathcal{N}_i$ as mentioned in Sec. \ref{sec:P_eff}. Hereafter, we assume that $\delta \mathcal{N}_i$ contains all the contributions.

Central to our calculation is the construction of the polarization operator. As we shall use a bond operator theory to calculate the optical conductivity (see next section), we find the following representation of the polarization operator very useful
\begin{equation}
 \mathbfcal{P}
 =\frac{1}{V} \sum_i e \delta \mathcal{N}_i {\bf r}_i
 =\frac{1}{V} \sum_{i,j} {\bf M}_{ij} \mathcal{I}_{ij},
 \label{eq:P_eff_symm}
\end{equation}
where we have written the polarization operator as the sum of the contributions coming from each bond $ij$ (each bond is counted only once). In the above expression we have
\begin{equation}
 {\bf M}_{ij} = M_{ij}^x \hat{x} + M_{ij}^y \hat{y} .
\end{equation}
The coefficients ${\bf M}_{ij}$ are constrained by the symmetries of the Hamiltonian (\ref{eq:H_eff_without_Gamma}): the lattice-translation, $C_6$ rotation and inversion. Once all independent ${\bf M}_{ij}$ are specified, the rest can be obtained by applying appropriate symmetry operators. For example, once we specify all ${\bf M}_{ij}$ within a unit cell, then the lattice-translation symmetry $T$ under which ${\mathcal{P}}$ does not change, implies
\begin{equation}
 {\bf M}_{i'j'}={\bf M}_{ij},
\end{equation}
where $i'=T(i)$, $j'=T(j)$ with $T$ generating the lattice-translation. Thus, it is sufficient to specify the form of ${\bf M}_{ij}$ within a unit cell of the lattice. If the unit cell contains more than one bond related by point group symmetries like $C_6$ rotation and Inversion, from the $C_6$ rotation we get:
\begin{equation}
 \left(
 \begin{array}{c}
 M_{i'j'}^x
 \\
 M_{i'j'}^y
 \end{array}
 \right)
 = 
 \left(
 \begin{array}{cc}
 \textup{cos}\frac{\pi}{3} & \textup{sin}\frac{\pi}{3}
 \\
 -\textup{sin}\frac{\pi}{3} & \textup{cos}\frac{\pi}{3}
 \end{array}
 \right) 
 \left(
 \begin{array}{c}
 M_{ij}^x
 \\
 M_{ij}^y
 \end{array}
 \right),
\end{equation}
where $C_6$ was taken as the counter clockwise rotation by $\pi/3$, and $i'=C_6(i)$, $j'=C_6(j)$. Similarly, from the Inversion $I$, we have
\begin{equation}
 {\bf M}_{i'j'}=-{\bf M}_{ij},
\end{equation}
where $i'=I(i)$, $j'=I(j)$. We shall use the above arguments to explicitly calculate the coefficients ${\bf M}_{ij}$ in different cases.

\section{Calculation of Optical conductivity in VBS state using bond operator mean-field theory\label{sec:bond_op_optical_conductivity}}

Having derived the simplified form (\ref{eq:P_eff_symm}) of the polarization operator that will be useful for VBS states (see below), we finally outline a general framework for the computation of the optical conductivity in VBS states by using bond operator mean-field theory. To this effect, we start with a brief description of the bond operator mean-field theory in presence of the DM interactions \cite{2012_Hwang}.
\subsection{Bond operator mean-field theory}
In the bond operator mean-field theory, we start with the bonds on which the dimers reside. The Hamiltonian on each bond is given by:
\begin{equation}
J_1 {\bf S}_L \cdot {\bf S}_R + D_1 \hat{z} \cdot {\bf S}_L \times {\bf S}_R
\end{equation}
where $J_1$ and $D_1$ denote the Heisenberg antiferromagnetic exchange and DM interactions, respectively, and ${\bf S}_L$ and ${\bf S}_R$ are the two spins participating in the dimer formation. The eigenstates of this Hamiltonian are given by
\begin{eqnarray}
\left| s \right> &=& \frac{1}{\sqrt{2}} ( e^{i \alpha/2} \left| \uparrow \downarrow \right> - e^{-i \alpha/2} \left| \downarrow \uparrow \right> ),
 \\
\left| t_{+} \right> &=& - \left| \uparrow \uparrow \right>,
 \\
\left| t_{0} \right> &=& \frac{1}{\sqrt{2}} ( e^{i \alpha/2} \left| \uparrow \downarrow \right> + e^{-i \alpha/2} \left| \downarrow \uparrow \right> ),
 \\
\left| t_{-} \right> &=& \left| \downarrow \downarrow \right>.
\label{eq:BOT_basis}
\end{eqnarray}
where $ e^{i \alpha} = \frac{J_1+iD_1}{\sqrt{J_1^2+D_1^2}}$.
Now we define {\it four} bond operators, $s,~t_{\tau}~(\tau=+,0,-)$, such that
\begin{equation}
 s^{\dagger} \left| 0 \right>= \left| s \right>,~ t_{+}^{\dagger} \left| 0 \right>= \left| t_{+} \right>,~t_{0}^{\dagger} \left| 0 \right> = \left| t_{0} \right>,~ t_{-}^{\dagger} \left| 0 \right> = \left| t_{-} \right>,
\end{equation}
where the bond operators satisfy the bosonic statistics and $\left| 0 \right>$ is the vacuum of the Fock space of the boson operators . While $s^\dagger$ creates the spin-singlet excitation, $\{t_+^{\dagger},t_0^{\dagger},t_-^{\dagger}\}$ create three spin-triplet excitations and hence constitute the triplon operators on the bond. The four states exhaust the spin Hilbert space on the bond and this translates into the hardcore constraint on the boson operators given by $s^\dagger s+\sum_{\tau=-1}^{1}t_{\tau}^\dagger t_{\tau}=1$. We can use the above bond operators to represent the two spin operators, ${\bf S}_L$ and ${\bf S}_R$  as

\begin{eqnarray}
 {S}_{L,\pm} &=& \frac{1}{\sqrt{2}} ( s^{\dagger} t_{\mp} + t_{\pm}^{\dagger} s \mp t_{\pm}^{\dagger} t_{0} \pm t_{0}^{\dagger} t_{\mp} ) \cdot e^{\mp i \alpha/2},\nonumber\\
 {S}_{L,z} &=& \frac{1}{2} ( s^{\dagger} t_{0} + t_{0}^{\dagger} s + t_{+}^{\dagger} t_{+} - t_{-}^{\dagger} t_{-} ),\nonumber\\
 {S}_{R,\pm} &=& \frac{1}{\sqrt{2}} ( - s^{\dagger} t_{\mp} - t_{\pm}^{\dagger} s \mp t_{\pm}^{\dagger} t_{0} \pm t_{0}^{\dagger} t_{\mp} ) \cdot e^{\pm i \alpha/2},\nonumber\\
 {S}_{R,z} &=& \frac{1}{2} ( - s^{\dagger} t_{0} - t_{0}^{\dagger} s + t_{+}^{\dagger} t_{+} - t_{-}^{\dagger} t_{-} ),
 \label{eq:bond_op}
\end{eqnarray}
where $S_{a,\pm}=S_{a,x} \pm iS_{a,y}$ $(a=L,R)$ as usual.

Using the bond operators, we can re-write the Hamiltonian (\ref{eq:H_eff_without_Gamma}) and impose the hardcore constraint through Lagrange multiplier $\mu$. With the spin Hamiltonian quartic in the bond operators, a mean-field description of the VBS-ordered ground state is obtained by condensing the $s$-bosons ({\it i.e.} $\bar{s} = \langle s \rangle \neq 0$) on the bonds, where dimers are present, and taking appropriate mean-field decouplings for remaining terms with the $t$-bosons. Both the singlet condensation amplitude $\bar{s}$ and Lagrange multiplier $\mu$ are then calculated self-consistently. The quadratic mean-field Hamiltonian for the spin-triplet excitations (triplons), then has the following form \cite{1990_Sachdev,1994_Gopalan,2012_Hwang}:
\begin{equation}
 \mathcal{H}_{quad} 
 =
 \frac{1}{2} \sum_{\tau=\pm,0} \sum_{\bf k} 
 \Lambda_{\tau}^{\dagger}({\bf k}) \mathcal{M}_{\tau}({\bf k}) \Lambda_{\tau}({\bf k}),
 \label{eq:H_quad}
\end{equation}
where
\begin{equation}
 \Lambda_{\tau}({\bf k}) = \left[ t_{\tau ,1}({\bf k}), \cdots ,t_{\tau ,\zeta}({\bf k}), t_{\bar{\tau} ,1}^{\dagger}(-{\bf k}), \cdots ,t_{\bar{\tau} ,\zeta}^{\dagger}(-{\bf k}) \right]^T,
 \nonumber
\end{equation}
$\bar{\tau}=-\tau$, $\zeta$ denotes the number of dimers within the unit cell, and $\mathcal{M}_{\tau}({\bf k})$ is the Hamiltonian matrix for the triplon excitations, which consists of hopping and pairing amplitudes of the $t$-bosons. The index $\tau~(=\pm 1,0)$ denotes the $z$-component of the spin quantum number, which is conserved because of the SO(2) spin-rotation symmetry of the Hamiltonian in Eq. (\ref{eq:H_eff_without_Gamma}). The above bosonic Hamiltonian in Eq. (\ref{eq:H_quad}) is diagonalized through the Bogoliubov transformation to yield the triplon spectra, $\omega_{\tau,b}({\bf k})~(b=1,\cdots,\zeta)$. 
For more details on the bond operator theory for anisotropic spin system, readers are referred to Ref. \cite{2012_Hwang}.

\subsection{Effective polarization and optical conductivity within bond operator mean-field theory}

We now express the polarization operator in terms of the bond operators by using the bond operator representations in Eq. (\ref{eq:bond_op}) for the the spin operators. The resulting expression of the polarization operator contains terms that are quartic in the $s$- and $t$-bosons. In the VBS ground state, the $s$-bosons are condensed and the polarization operator has nonzero matrix elements for the transitions from the VBS ground state to triplon-excited states. In other words, when the system is placed in an external spatially uniform a.c. electric field, it generates triplon excitations. We find that we can write the polarization operator as
\begin{equation}
 \mathbfcal{P} = \mathbfcal{P}^{(2)}+\mathbfcal{P}^{(3)}+\mathbfcal{P}^{(4)}
\end{equation}
where $\mathbfcal{P}^{(n)}$ denotes the contribution from terms containing $n$-triplon excitations. 
The absence of $\mathbfcal{P}^{(0)}$ term suggests that there is no polarization response in the ground state. 
This is due to the fact that there is no quadratic term in $s$ operator in Eq. (\ref{eq:bond_op}).

Since the triplon excitations are gapped, at low frequencies, we expect that the matrix elements corresponding to creations of the minimum number of  triplons dominate the optical response. Hence, we only consider the quadratic part $\mathbfcal{P}^{(2)}$, which is concerned with two-triplon excitations from the ground state. Then, the matrix element of the polarization operator is given by
\label{eq:quadratic_P_matrix_element}
\begin{equation}
 \langle \psi_n | \mathbfcal{P}^{(2)} | \psi_0 \rangle 
 =
 \frac{1}{V}
 \sum_{ij} {\bf M}_{ij} \langle \psi_n | \mathcal{I}_{ij}^{(2)} | \psi_0 \rangle,
\end{equation}
where $\mathcal{I}_{ij}^{(2)}$ is the quadratic part of $\mathcal{I}_{ij}$ in terms of the $t$-bosons.
Plugging this into Eq. (\ref{eq:Re_sigma_final}), we have the following expression for the optical conductivity:
\begin{equation}
 \sigma^{(2)} (\omega)
 =
 \frac{\pi\omega}{\hbar V} \sum_{n\ne0} 
 \left| \sum_{ij} M_{ij}^{x,y} \langle \psi_n | \mathcal{I}_{ij}^{(2)} | \psi_0 \rangle \right|^2
 \delta (\omega - \omega_{n}).
 \label{eq:sigma2}
\end{equation}
In this expression, the excited states $| \psi_n \rangle$ are constrained to the two-triplon states with the zero momentum and zero spin $z$-component since the ground state $| \psi_0 \rangle$ has zero value for the momentum and spin $z$-component and the polarization operator $\mathbfcal{P}$ preserves both quantities. Consequently, $\omega_n$ is the associated two-triplon excitation energy.

\section{Optical conductivity in deformed Kagome lattice antiferromagnet Rb$_2$Cu$_3$SnF$_{12}$ \label{sec:deformed_kagome_lattice}}
\begin{figure}
 \centering
 \includegraphics[width=0.5\linewidth]{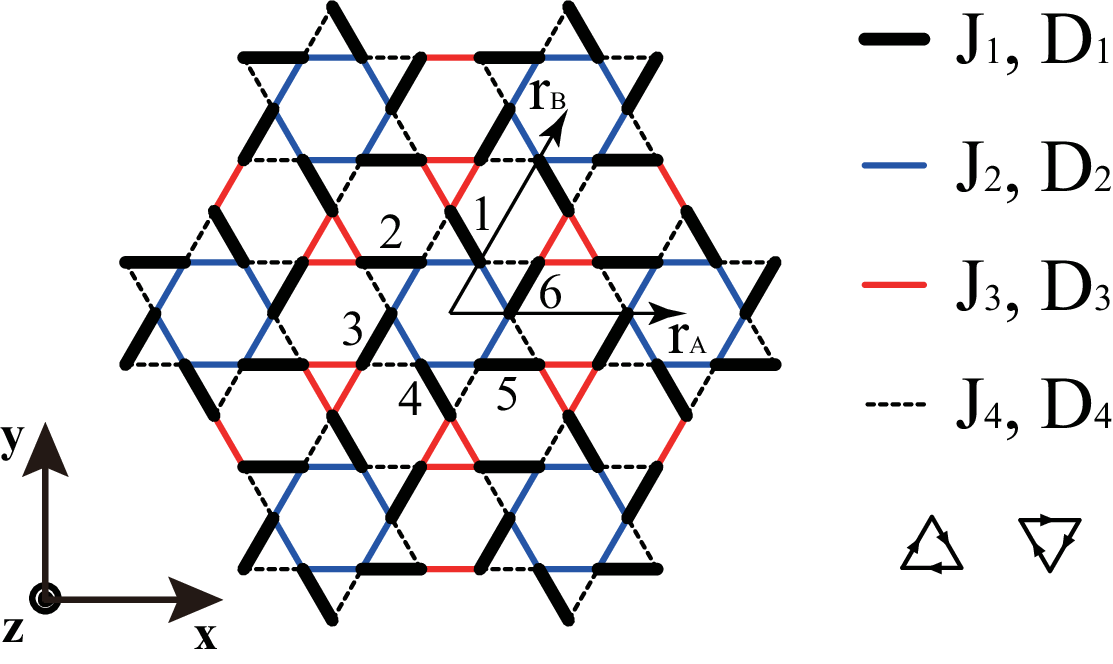}
 \caption{(Color online) The deformed Kagome lattice of Rb$_2$Cu$_3$SnF$_{12}$ with four different couplings for the Heisenberg and Dzyaloshinskii-Moriya interactions respectively. The DM vectors are assumed to be perpendicular to the lattice plane and their direction is along the positive direction of the z-axis when the orientation from $i$ to $j$ in $\mathrm{\bf S}_i \times \mathrm{\bf S}_j$ is clockwise as indicated by the arrows in the lower right triangles. For the VBS state with the 12-site unit cell, the valence bonds are formed in the interaction links with the strongest couplings, $J_1$ and $D_1$ (thick black).
 \label{fig:deformed_kagome_lattice}}
\end{figure}
Having derived the form of the optical conductivity, we now show that measurements of the optical conductivity can yield useful information about VBS ordered ground states in frustrated magnets. We first focus on the compound Rb$_2$Cu$_3$SnF$_{12}$, which is a spin-1/2 antiferromagnet on a deformed Kagome lattice. Before sketching out our calculation, we give a brief description of the compound. The system has 12-site unit cell as shown in Fig. \ref{fig:deformed_kagome_lattice}. The lattice vectors are given by
\begin{equation}
 {\bf r}_A = 4 a \hat{x}, 
 ~ 
 {\bf r}_B = 4 a \left( \frac{1}{2}\hat{x} + \frac{\sqrt{3}}{2}\hat{y} \right), 
 ~ 
 {\bf r}_C = {\bf r}_B - {\bf r}_A,
 \nonumber
\end{equation}
where $a$ is the lattice spacing in the Kagome lattice. Due to lattice deformation from the ideal Kagome lattice structure, there are four different types of couplings for $J_{ij}$ and ${\bf D}_{ij}$, respectively (see Fig. \ref{fig:deformed_kagome_lattice}).

The ground state of Rb$_2$Cu$_3$SnF$_{12}$ has the 12-site pinwheel valence bond solid (VBS) with  dimers at the strongest $J_1$-links in Fig. \ref{fig:deformed_kagome_lattice}. According to a combined study of neutron scattering experiment and dimer series expansion \cite{2010_Matan}, the antiferromagnet is described by the Hamiltonian (\ref{eq:H_eff_final_form}) in two dimensions where the last anisotropic term $(\propto \Gamma^{ab})$ is expected to be small and hence can be ignored. Further, the directions of the DM vectors are, to a good approximation, perpendicular to the Kagome plane, {\it i.e.}, ${\bf D}_{ij}=D_{ij} \hat{z}$, which is consistent with our approximations. Hence, the minimal spin-Hamiltonian for Rb$_2$Cu$_3$SnF$_{12}$ is given by Eq. (\ref{eq:H_eff_without_Gamma}) with the four different coupling constants obtained in a previous study \cite{2010_Matan}:
\begin{eqnarray}
 J_1=18.6 ~ \textup{meV},
 ~
 J_2/J_1=0.95,
 ~
 J_3/J_1=0.85,
 ~
 J_4/J_1=0.55,
 \nonumber\\
 D_n/J_n=0.18 ~(n=1,\cdots,4).
 \label{eq:deformed_kagome_parameters}
\end{eqnarray} 

Despite the above anisotropy, the Hamiltonian (\ref{eq:H_eff_without_Gamma}) has all the symmetries  mentioned in Sec. \ref{sec:kagome_optical_conductivity}.

\subsection{Triplon excitation spectra}

\begin{figure*}
 \centering
 \includegraphics[width=0.45\linewidth,angle=270]{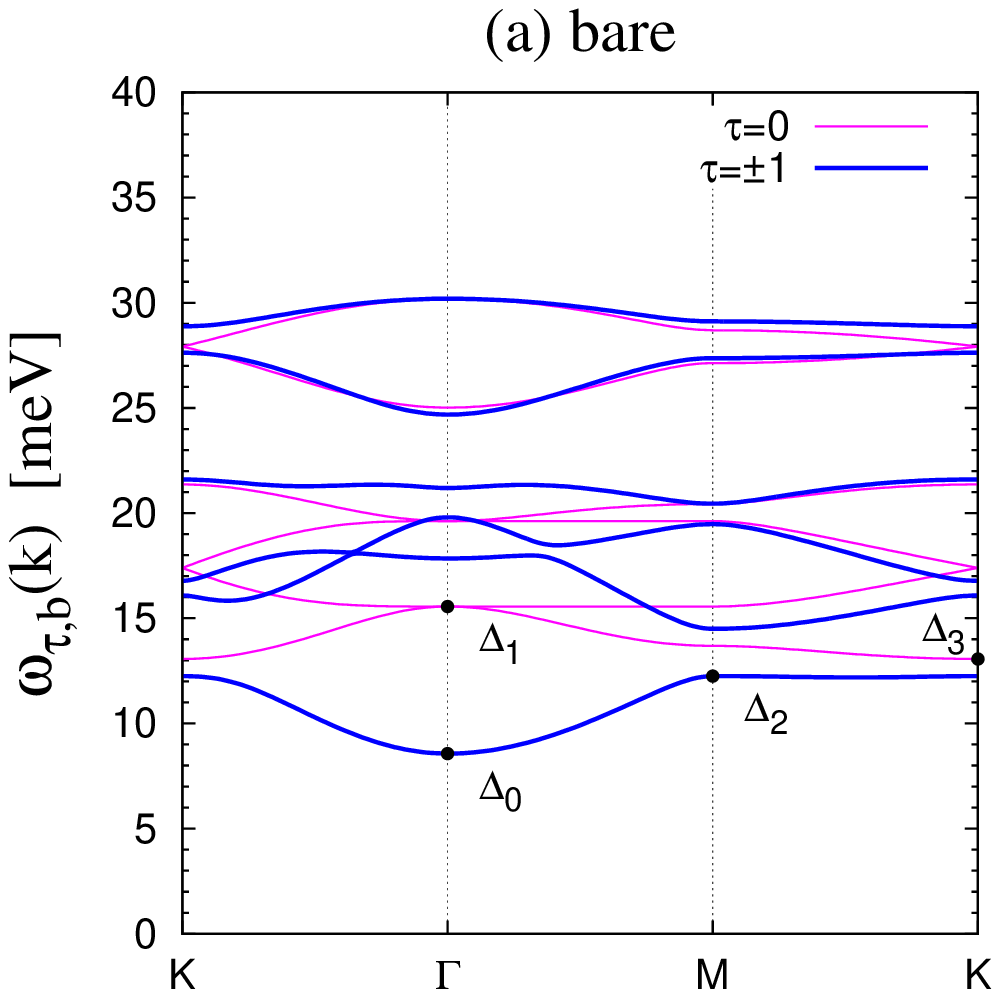}
 ~~~~~~~~~~
 \includegraphics[width=0.45\linewidth,angle=270]{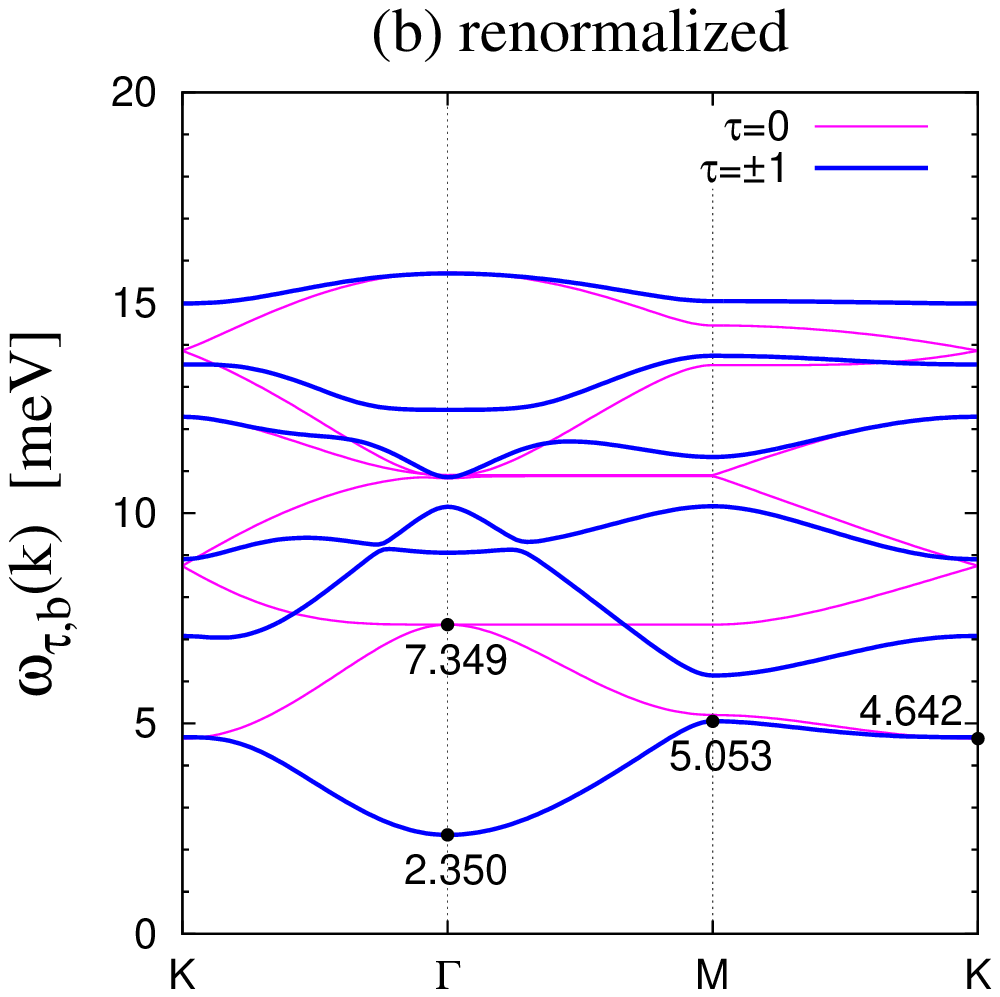}
 \caption{(Color online) Single-triplon excitation spectra $\omega_{\tau,b}({\bf k}) ~ (\tau=+,0,-;b=1,\cdots,6)$ from the bond operator theory.
 (a) Bare triplon excitations calculated with the coupling constants in (\ref{eq:deformed_kagome_parameters}). 
 (b) Renormalized triplon excitations calculated with the effective coupling constants in (\ref{eq:deformed_kagome_fitting_parameters}).
 The renormalized triplon excitations are obtained by adjusting the exchange coupling constants for the characteristic energies $\Delta_n~(n=0,\cdots,3)$ to have correct energy scales found in the neutron scattering and series expansion (see \ref{appendix:fitting}).
 \label{fig:triplon_dispersion}}
\end{figure*}

In the VBS phase, the elementary triplon excitations are obtained by breaking the spin-singlet bonds into the spin-triplet states. A previous study by two of the present authors \cite{2012_Hwang}, on the above deformed Kagome lattice antiferromagnet, investigated the triplon excitations in the valence bond solid state by employing the bond operator mean-field theory for the spin model Eq. (\ref{eq:H_eff_without_Gamma}).   
In Ref. \cite{2012_Hwang}, the triplon excitation spectra were obtained from the above bond operator theory using the parameters in Eq. (\ref{eq:deformed_kagome_parameters}). Although it captures qualitative features in the neutron scattering results such as the position of the lowest energy gap for the triplon excitations and the band curvatures around high symmetry points, the energy scales are inconsistent with the experimental results \cite{2010_Matan}. This discrepancy is due to the possible renormalizations of the mean-field parameters owing to fluctuations and effects of higher order terms beyond the quadratic mean-field theory. A quantitative match of the triplon spectra may be obtained once the renormalizations are taken into account. Here, we achieve that phenomenologically by fitting the experimental triplon spectra by varying the exchange couplings. The details of the fitting are given in \ref{appendix:fitting}. The renormalized coupling constants are
\begin{equation}
 \begin{array}{lrrrlrrr}
 \tilde{J}_1 &=&  7.730 & \textup{meV}, & \tilde{D}_1 &=& -0.245 & \textup{meV},
 \\
 \tilde{J}_2 &=& 9.400 & \textup{meV}, & \tilde{D}_2 &=& 3.285 & \textup{meV},
 \\
 \tilde{J}_3 &=&  6.976 & \textup{meV}, & \tilde{D}_3 &=&  0.772 & \textup{meV},
 \\ 
 \tilde{J}_4 &=& 2.296 & \textup{meV}, & \tilde{D}_4 &=& 2.179 & \textup{meV}.
 \end{array}
 \label{eq:deformed_kagome_fitting_parameters}
\end{equation}
The corresponding triplon spetra are shown in Fig. \ref{fig:triplon_dispersion}. 
While the above fitting is somewhat ad-hoc, it restores the quantitative features of the low energy part of the triplon spectra.

\subsection{Optical conductivity}

\begin{figure}
 \centering
 \includegraphics[scale=0.5]{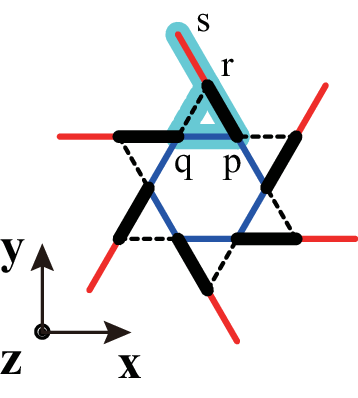}
 \caption{(Color online) 
 Interaction bonds in a unit cell of the deformed Kagome lattice. 
 The thick, light blue line denotes four independent bonds not related by symmetries of the system.
 Four spins in the bonds are labelled by $p,q,r,s$.
 \label{fig:deformed_kagome_lattice_unit_cell}}
\end{figure}

For the optical conductivity computation, we find $\mathbfcal{P}$ in Eq. (\ref{eq:P_eff_symm}) for the 12-site unit cell of the pinwheel VBS phase. As mentioned in Sec. \ref{sec:kagome_optical_conductivity}, we determine the coefficient ${\bf M}_{ij}$ for independent bonds in a unit cell and generate other coefficients by applying symmetry operations of the system. In the 12-site unit cell of the pinwheel VBS state, there are four independent bonds that are not related by symmetries. These four bonds are marked with thick, light blue line in Fig. \ref{fig:deformed_kagome_lattice_unit_cell} with four spins in the bonds being labelled with $p,q,r,s$. For the four bonds, the coefficients $\{ {\bf M}_{ij} \}$ are explicitly calculated to be
\begin{eqnarray}
 {\bf M}_{pq}= A \frac{2}{\sqrt{3}} \hat{y},~~
 {\bf M}_{qr}= A \left( \hat{x} - \hat{y} \frac{1}{\sqrt{3}} \right),
 \nonumber\\
 {\bf M}_{rp}= A \left( - \hat{x} - \hat{y} \frac{1}{\sqrt{3}} \right),~~
 {\bf M}_{rs}= B \left( \hat{x} + \hat{y} \frac{1}{\sqrt{3}} \right),
\end{eqnarray}
where
\begin{equation}
 A = \sqrt{\frac{J_1J_2J_4}{U^3}} \cdot \frac{3ea}{2},
 ~~~
 B = \sqrt{\frac{J_3^3}{U^3}} \cdot \frac{3ea}{2}.
\end{equation}
All the other bonds in the unit cell are related by six-fold rotational symmetry operation $C_6$ and can be generated by repeated applications of $C_6$. We tabulate them in \ref{appen_table} (Table \ref{tab:L}) for completeness. We can then calculate the optical conductivity using Eq. (\ref{eq:sigma2}). We present below the results of the optical conductivity for the both cases: (i) with the bare coupling constants in Eq. (\ref{eq:deformed_kagome_parameters}) and (ii) with the renormalized coupling constants in Eq. (\ref{eq:deformed_kagome_fitting_parameters}).

\begin{figure*}
 \centering
 \includegraphics[width=0.6\linewidth,angle=270]{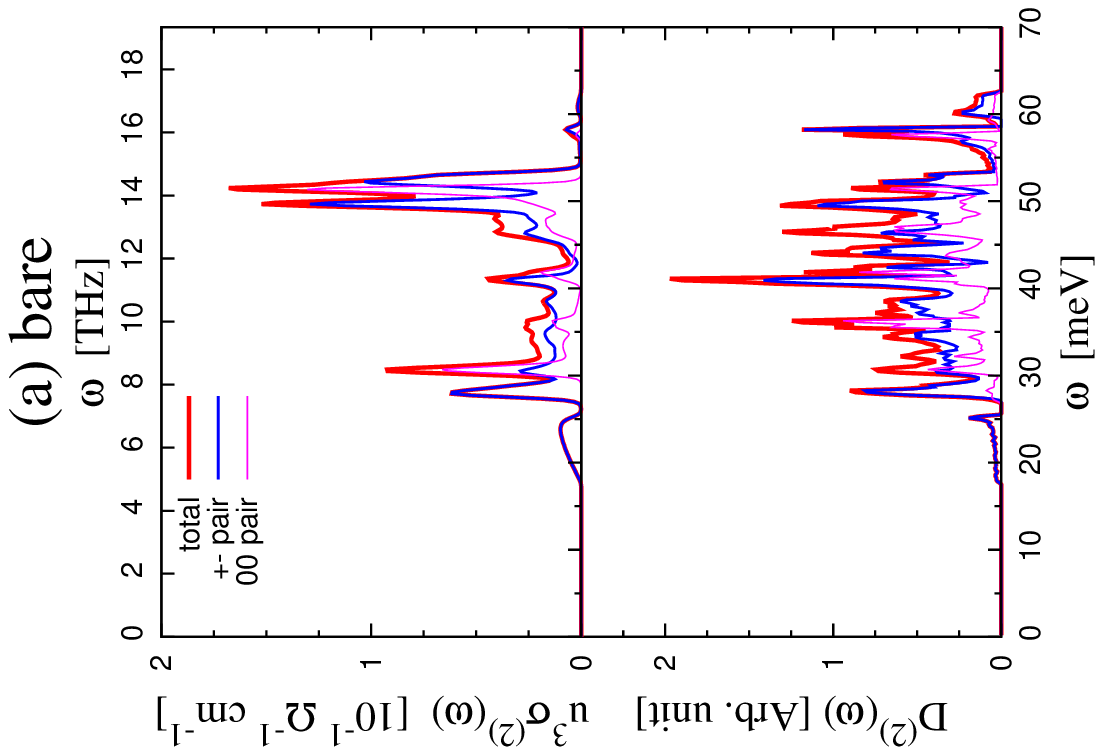}
 ~~~
 \includegraphics[width=0.6\linewidth,angle=270]{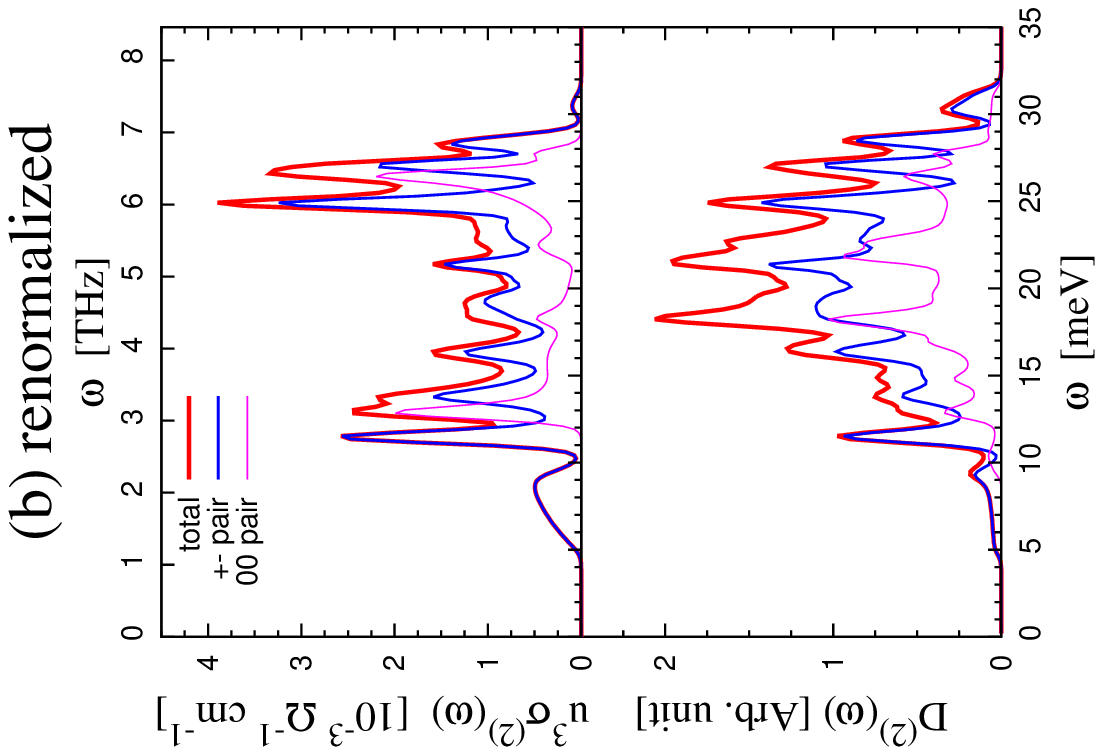}
 \caption{(Color online) Optical conductivity, $\sigma^{(2)}({\omega})$, and density of states, $D^{(2)}({\omega})$, of the deformed kagome lattice antiferromagnet Rb$_2$Cu$_3$SnF$_{12}$. (a) The results with the bare coupling constants in Eq. (\ref{eq:deformed_kagome_parameters}). (b) The results with the renormalized coupling constants in Eq. (\ref{eq:deformed_kagome_fitting_parameters}). For the optical conductivity, $u^3 \cdot \sigma^{(2)}(\omega)$ is plotted, where $u$ is the Coulomb repulsion energy scale in the unit of $eV$. In the above plots, $\sigma^{(2)}({\omega})$ and $D^{(2)}({\omega})$ are decomposed into $+-$ and $00$ components  according to the spin structure of the two-triplon excitation: $| +, b_1, {\bf Q}; -, b_2, -{\bf Q} \rangle$ (blue lines) and $| 0, b_1, {\bf Q}; 0, b_2, -{\bf Q} \rangle$ (magenta lines). The sum of the two components is denoted with red lines in the figure. In the calculation, the delta functions in $\sigma^{(2)}(\omega)$ and $D^{(2)}(\omega)$ were replaced with the smooth function $\delta_{\mathcal{T}} (\omega)=-\frac{\partial}{\partial \omega} \frac{1}{e^{\omega/ \mathcal{T}}+1}$ with $\mathcal{T}=0.01J_1\simeq0.2~\textup{meV}$.
 \label{fig:conductivity_deformed_kagome}}
\end{figure*}
\subsection{Result and discussion\label{sec:deformed_kagome_optical_conductivity}}

Figure \ref{fig:conductivity_deformed_kagome} shows the  results of our calculation of the two-triplon contribution to the optical conductivity, $\sigma^{(2)}(\omega)$, for the deformed Kagome lattice antiferromagnet Rb$_2$Cu$_3$SnF$_{12}$. The left and right panels in the figure show the results calculated with the bare and renormalized coupling constants, respectively. We may have a rough estimation on the order of magnitudes of the optical conductivity by computing the dimensionful prefactor in Eq. (\ref{eq:sigma2}) given by:
\begin{equation}
 \frac{\pi B^2}{\hbar V} 
 =
 N_{uc}^{-1}
 \frac{9\sqrt{3}}{32} \pi \left( \frac{J_3}{U} \right)^3 \frac{e^2}{\hbar c}
 \nonumber\\
 =
 u^{-3} N_{uc}^{-1} 
 \times
 \left\{
 \begin{array}{c}
 7.232 \times 10^{-3}   
 \\
 6.213 \times 10^{-4} 
 \end{array}
 \right\}
 \Omega^{-1}\textup{cm}^{-1}
 ,
 \label{eq:sigam2-prefactor}
\end{equation}
where $N_{uc}$ is the number of unit cells, $c$ (=20.356\AA) is the size of the unit cell along the $z$ axis in the compound \cite{2008_Morita}, and we have taken into account the fact that there are three Kagome layers in a unit cell of Rb$_2$Cu$_3$SnF$_{12}$. In the last equality of the above equation, the upper and lower values are computed with the bare and renormalized coupling constants, respectively. For the Coulomb repulsion energy, we set $U=u ~\textup{eV}$ and consider $u$ in the range 6$\sim$9 \cite{2006_Nakamura,2013_Jeschke}. However, since it is very difficult to determine the accurate value of $U$ in the compound, we plot $u^3 \cdot \sigma^{(2)}(\omega)$ for the optical conductivity in Fig. \ref{fig:conductivity_deformed_kagome}. According to our calculations with the bare and renormalized coupling constants, the optical conductivity $\sigma^{(2)}(\omega)$ of the antiferromagnet Rb$_2$Cu$_3$SnF$_{12}$ is in the order of $10^{-6} \sim 10^{-4}~\Omega^{-1}\textup{cm}^{-1}$  for $u=6\sim9$. As explained in the previous section, the two-triplon excitations with the zero momentum and zero spin $z$-component for the intermediate states contribute finite matrix elements for the optical conductivity computation. The density of the states, $D^{(2)}(\omega)$, for such two-triplon excitations is plotted in Fig. \ref{fig:conductivity_deformed_kagome} as well:
\begin{equation}
 D^{(2)}(\omega)=\frac{1}{V}\sum_{n \ne 0}\delta(\omega-\omega_{n}),
\end{equation}
where $\omega_{n}$ is the two-triplon excitation energy.

In Fig. \ref{fig:conductivity_deformed_kagome}, $\sigma^{(2)}(\omega)$ and $D^{(2)}(\omega)$ are decomposed into $+-$ and $00$ components according to the spin structure of the two-triplon intermediate states: $| +, b_1, {\bf Q}; -, b_2, -{\bf Q} \rangle$ (blue lines) and $| 0, b_1, {\bf Q}; 0, b_2, -{\bf Q} \rangle$ (magenta lines), where $b_{1,2}$ and $\pm{\bf Q}$ denote the band indices and triplon momenta, respectively. The sum of the two contributions is denoted with red lines in the figure.

Each contribution to the optical conductivity consists of several peaks with various magnitudes and widths. Those peaks originate from the high density of states of the two-triplon excitations as can be seen from the similarity of the peak structures in $\sigma^{(2)}(\omega)$ and $D^{(2)}(\omega)$. This is expected because of the presence of the delta-function factor, $\delta(\omega-\omega_n)$, in the expression of $\sigma^{(2)}(\omega)$ in Eq. (\ref{eq:sigma2}). Hence, we expect, apart from subtle cancellation due to the matrix elements, both of them to be correlated. So, the optical conductivity signal $\sigma^{(2)}(\omega)$ can provide useful information on the two-tripon excitations with zero-momentum and zero-spin in the antiferromagnet Rb$_2$Cu$_3$SnF$_{12}$.

Comparing the optical conductivities in Fig. \ref{fig:conductivity_deformed_kagome} (a) and (b), we notice that they show similar overall behaviors, particularly with almost the same pattern in the low frequency part. It is found in both cases that a broad hump at lowest frequencies is followed by a large and narrow peak. Looking at the corresponding density of states, we also notice resemblance between the two cases. This resemblance comes from the fact that both cases have the qualitatively the same behavior in low energy triplon dispersions. Beyond the low frequency part, they are similar to each other in overall shape with differences in peak positions and sizes. We discuss below about more details in the optical conductivity focusing on the low frequency part where the both cases show the same behavior.

In the low frequency region, the optical conductivity signal in Fig. \ref{fig:conductivity_deformed_kagome} (b) has two distinct peaks: a broad peak between 5 and 10 meV and a relatively sharp peak around 11$\sim$12 meV. Those peaks are solely generated by exciting the two-triplon states $\{ | +, b_1, {\bf Q}; -, b_2, -{\bf Q} \rangle \}$. Besides the above excitations, another peak that contributes to the low-energy part of the two-triplon density of states comes from the excitations which are related to the states $\{ | 0, 1, {\bf Q}; 0, 1, -{\bf Q} \rangle \}$, as shown in the plot of $D^{(2)}(\omega)$ (magenta) in Fig. \ref{fig:conductivity_deformed_kagome}. However, these states do not contribute to $\sigma^{(2)}(\omega)$ due to cancellation among the transition amplitudes for the excited states. The same pattern is also observed in the corresponding low frequency part of Fig. \ref{fig:conductivity_deformed_kagome} (a).

\begin{figure*}
 \centering
 \includegraphics[width=0.9\linewidth]{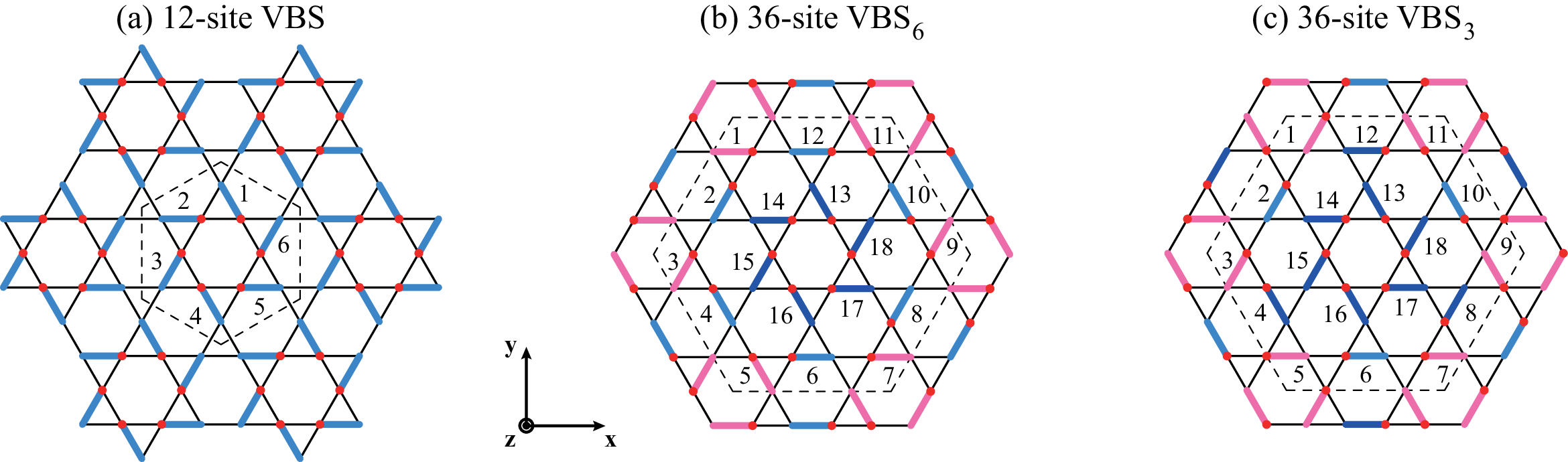}
 \caption{(Color online) Three possible valence bond solid states in the ideal Kagome lattice antiferromagnet.
 (a) 12-site VBS.
 (b) 36-site VBS$_6$.
 (c) 36-site VBS$_3$.
 The first number in the labelling implies the number of sites in a unit cell, and the subscripts mean their lattice rotation symmetries.
 The 12-site VBS and 36-site VBS$_6$ have sixfold rotation and inversion symmetries and 36-site VBS$_3$ has only threefold rotation symmetry.
 Thick line segments in colors denote dimer configuration for each VBS state.
 Numbers in the figure are dimer indices in a unit cell denoted with dashed line.
 In (b) and (c), three different colors (magenta, blue, and light blue) for dimers indicate the three distinct groups of dimers that are not related with symmetry operations.
 Red dots in the figure denote our convention for the position of the right spin ${\bf S}_R$ in bond operator representation at each dimer.
 \label{fig:VBS_patterns}}
\end{figure*}

\section{Optical conductivity calculation for the ideal Kagome lattice antiferromagnet\label{sec:ideal_kagome_lattice}}

In this section, we consider possible valence bond solid states in an ideal Kagome lattice antiferromagnet and investigate their low frequency optical responses. For the ideal Kagome lattice antiferromagnet, the Hamiltonian is given by (\ref{eq:H_eff_without_Gamma}) with uniform coupling constants $J_{ij}=J$ and $D_{ij}=D$ for nearest neighbors and the same orientations for the DM vectors as denoted in Fig. \ref{fig:deformed_kagome_lattice}:
\begin{equation}
\mathcal{H} 
=
\sum_{\langle i,j \rangle} 
\left(
J {\bf S}_i \cdot {\bf S}_j 
+ 
{D} \hat{z} \cdot {\bf S}_i \times {\bf S}_j 
\right).
\label{eq:H_ideal_kagome}
\end{equation}
In the $D=0$ limit, several possible VBS orders have been proposed. Among them, a 12-site VBS state and two 36-site VBS states (denoted by VBS$_6$ and VBS$_3$ respectively) are prominent \cite{2011_Huh,1991_Marston,2003_Nikolic,2007_Singh}. These VBS ordering patterns are shown in Fig. \ref{fig:VBS_patterns}. Here, the VBS$_6$ and VBS$_3$ phases have six-fold and three-fold rotation symmetries about the center of the VBS unit cell (dashed line), respectively. While the ground state of the antiferromagnet on ideal Kagome lattice is still controversial, recent numerical calculations suggest that these VBS states can be stabilized by tuning very small next nearest neighbour Heisenberg coupling \cite{2011_Iqbal}. For small but finite $D$, we expect these VBS phases to survive since they are gapped. At the mean field level of our calculation, the role of the next nearest neighbour exchanges is quantitative renormalization of the mean-field parameters. Though these next nearest neighbour exchanges seems to be absolutely crucial for the actual stabilization of the VBS order in the microscopic model, for the mean field theory of the VBS phase they provide qualitative changes. Hence we can ignore them in our present calculation.

Appropriately generalizing our bond operator theory to the three cases, we calculate the triplon excitations and from them we calculate optical conductivity. Unlike the deformed case, we only need to specify ${\bf M}_{ij}$ for one bond and all other such coefficients can be obtained by application of the space group symmetries of the spin model (\ref{eq:H_ideal_kagome}) on the ideal Kagome lattice. Here we discuss the results of our calculations.

\paragraph{12-site VBS:}

\begin{figure*}
 \centering
 \includegraphics[width=0.7\linewidth,angle=270]{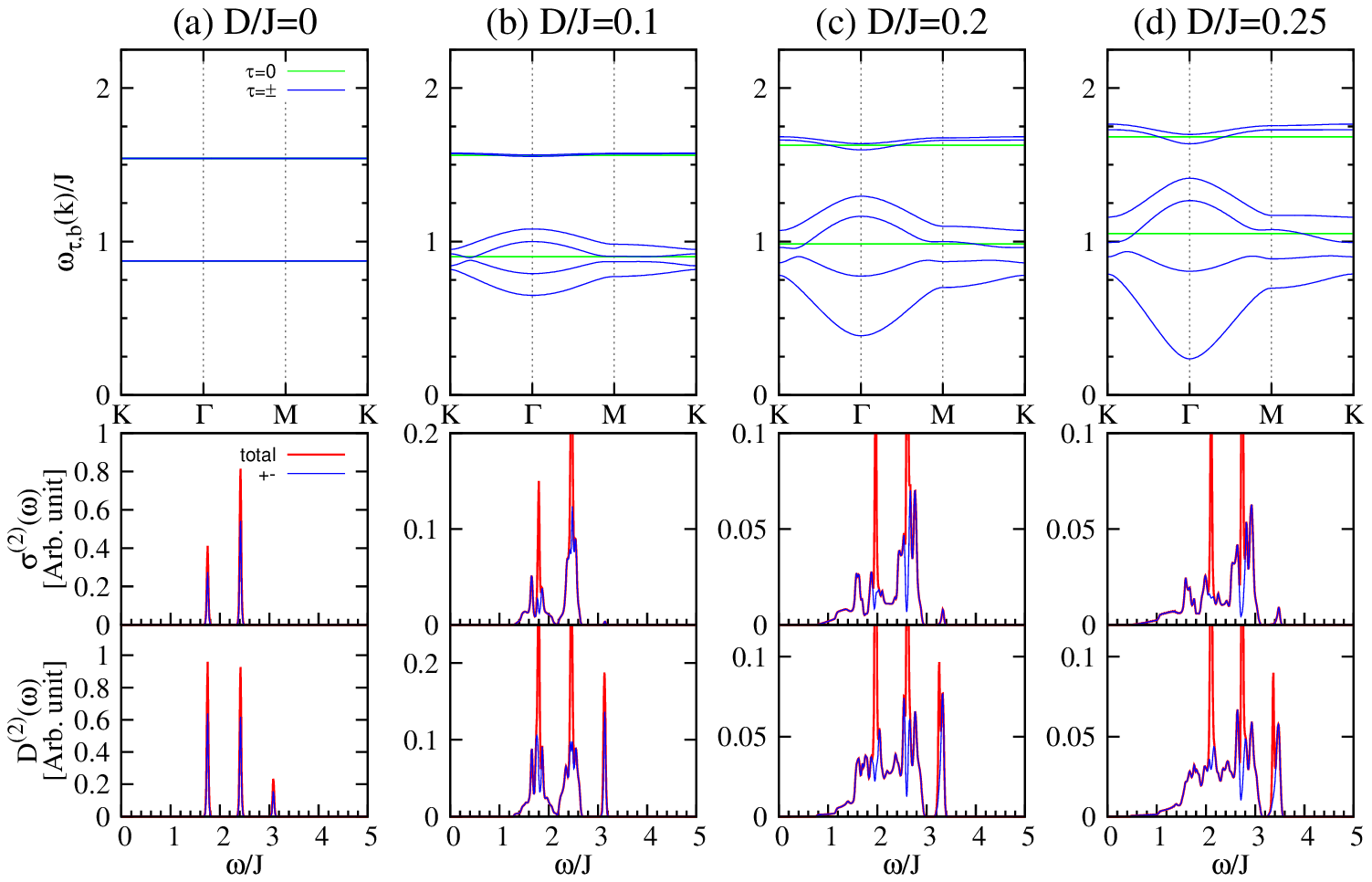}
 \caption{(Color online) Single-triplon dispersion $\omega_{\tau,b}({\bf k})$ ($\tau=+,0,-,~b=1,\cdots,6$), optical conductivity $\sigma^{(2)}({\omega})$, and density of states $D^{(2)}(\omega)$ of the 12-site VBS state.  Each column shows $\omega_{\tau,b}({\bf k})$, $\sigma^{(2)}({\omega})$, and $D^{(2)}(\omega)$ for a given value of $D/J$.  In the plots of the triplon dispersions, degenerate $\tau=\pm$ dispersions are drawn with blue lines and the other $\tau=0$ dispersion with green lines.  When $D/J=0$, the blue and green lines are overlapped due to the fact that the $\tau=+,0,-$ dispersions are all degenerate.  In the plots of the optical conductivity and density of states,  we only plot the $+-$ component (blue) and total sum (red) of the $+-$ and 00 components to avoid confusion. \label{fig:12_VBS_band_conductivity}}
\end{figure*}

The dimer pattern in the 12-site VBS state on the ideal Kagome lattice is essentially the same state that we considered for the deformed Kagome lattice antiferromagnet.
Despite the same VBS pattern, triplon excitations in both cases are quite different due to the difference in their coupling constants $\{J_{ij},~D_{ij}\}$, as shown in Fig. \ref{fig:12_VBS_band_conductivity}. A distinct feature in the present case is the presence of completely flat triplon bands in $D=0$ limit which arises from the so called {\it topologically orthogonal} dimer structures in the 12-site VBS on the ideal Kagome lattice \cite{2008_Yang,2011_Hwang}.

The flat triplon dispersions generate three distinct delta-function peaks in the density of states $D^{(2)}(\omega)$ of two-triplon excitations as shown in Fig. \ref{fig:12_VBS_band_conductivity} (a). These correspond to delta-function peaks in the optical conductivity signal $\sigma^{(2)}(\omega)$ at the same energies with appropriate weights on the peak sizes.
The peak with the highest energy does not appear in $\sigma^{(2)}$ 
since the transition amplitudes to the corresponding two-triplon excitations cancel each other.

Upon increasing the strength of the DM interaction, the triplon energy bands with $\tau=\pm$ become dispersive more and more whereas the bands with $\tau=0$ remain flat (see Fig. \ref{fig:12_VBS_band_conductivity} (b) $\sim$ (d)) since the DM interaction generates the hopping and pairing amplitudes only for the $t_{\tau=\pm}$ bosons. The $\tau=\pm$ triplon bands are degenerate due to the time reversal and inversion symmetries. With the increase of the triplon dispersion, as expected, the delta-function peaks in the two triplon density of states smear out and broaden and this is directly reflected in the optical conductivity response.

\paragraph{36-site VBS$_6$:}

\begin{figure*}
 \centering
 \includegraphics[width=0.7\linewidth,angle=270]{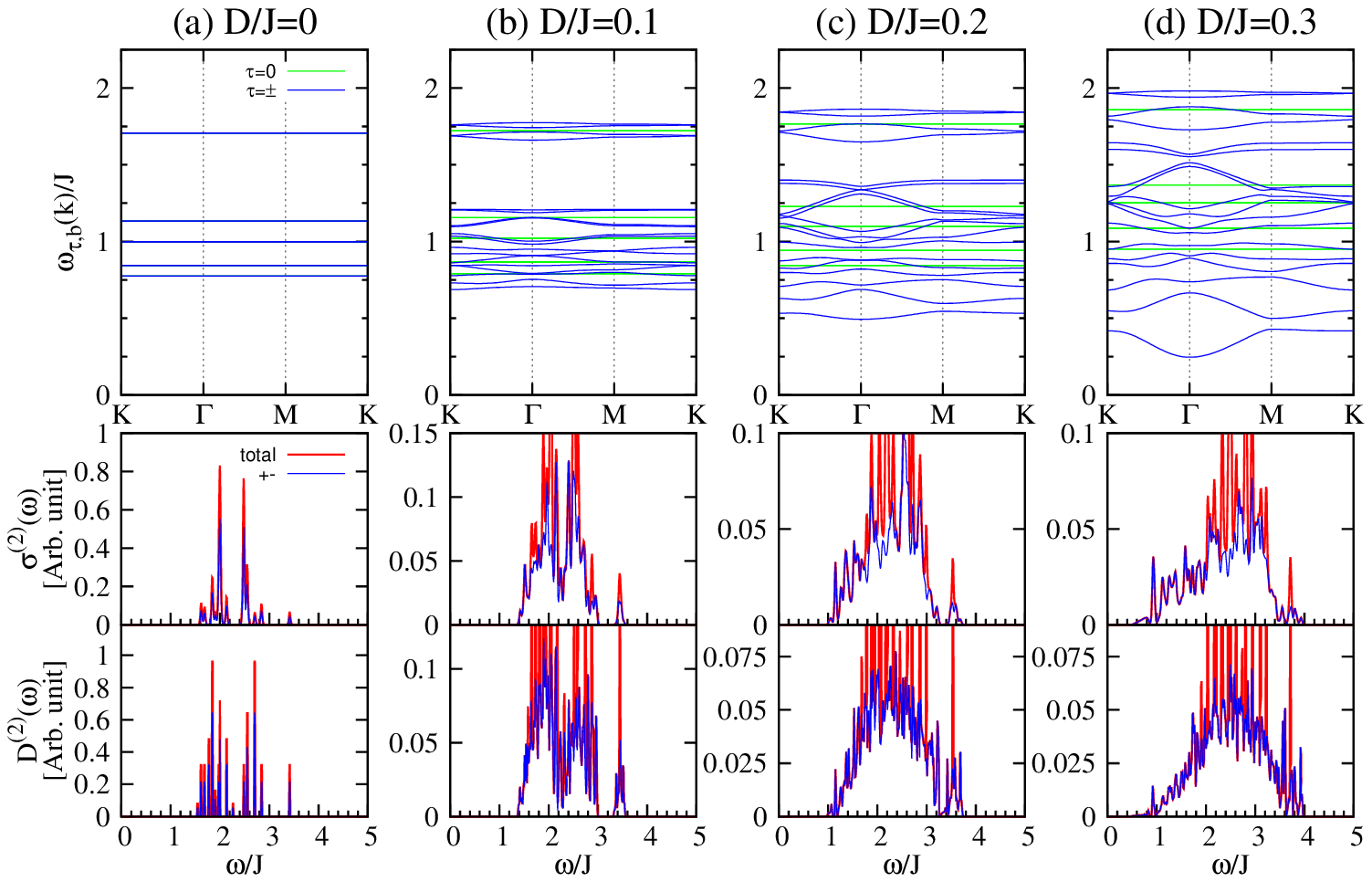}
 \caption{(Color online) Single-triplon dispersion $\omega_{\tau,b}({\bf k})$ ($\tau=+,0,-,~b=1,\cdots,18$), optical conductivity $\sigma^{(2)}({\omega})$, and density of states $D^{(2)}(\omega)$ of the 36-site VBS$_6$ state.
 Each column shows $\omega_{\tau,b}({\bf k})$, $\sigma^{(2)}({\omega})$, and $D^{(2)}(\omega)$ for a given value of $D/J$.
 In the plots of the triplon dispersions, degenerate $\tau=\pm$ dispersions are drawn with blue lines and the other $\tau=0$ dispersion with green lines.
 When $D/J=0$, the blue and green lines are overlapped due to the fact that the $\tau=+,0,-$ dispersions are all degenerate.
 In the plots of the optical conductivity and density of states,
 we only plot the $+-$ component (blue) and total sum (red) of the $+-$ and 00 components to avoid confusion.
 \label{fig:36_VBS_6_band_conductivity}}
\end{figure*}

In the case of the 36-site VBS$_6$ state, the situation is quite similar to that of the 12-site VBS state. The 36-site VBS$_6$ in Fig. \ref{fig:VBS_patterns} (b) has the topologically orthogonal dimer structures. These structures lead to decoupling of the dimers in the VBS state into independent dimers and clusters (when $D/J=0$). The six independent dimers (13 $\sim$ 18 in Fig. \ref{fig:VBS_patterns} (b)) contribute to the flat triplon energy band at the excitation energy of $J$ with sixfold degeneracy. The other flat bands come from two six-dimer clusters. As before, the flat bands lead to several isolated delta peaks in $D^{(2)}(\omega)$ and consequently in $\sigma^{(2)}(\omega)$. Under nonzero DM interactions, the degenerate flat triplon bands in the $\tau=\pm$ sectors get dispersive, with the lowest triplon energy gap at the K point ($0 < D/J < 1.2$) and the $\Gamma$ point ($D/J \geq 1.2$) in the first Brillouin zone, and the corresponding delta peaks in $D^{(2)}$ and $\sigma^{(2)}$ spread and merge into relatively small and broad signals as shown in Fig. \ref{fig:36_VBS_6_band_conductivity}.

\paragraph{36-site VBS$_3$:}

\begin{figure*}
 \centering
 \includegraphics[width=0.7\linewidth,angle=270]{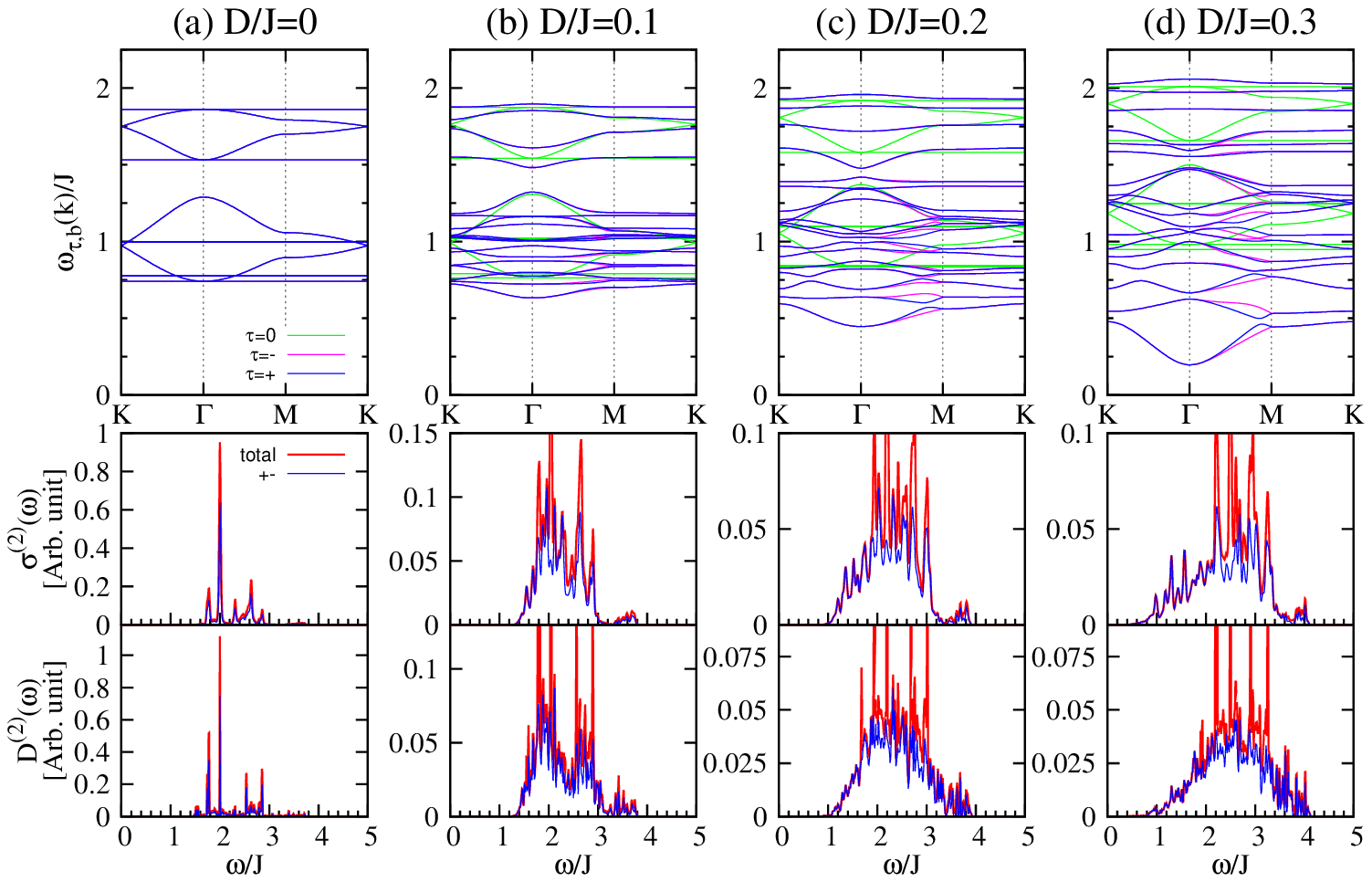}
 \caption{(Color online) Single-triplon dispersion $\omega_{\tau,b}({\bf k})$ ($\tau=+,0,-,~b=1,\cdots,18$), optical conductivity $\sigma^{(2)}({\omega})$, and density of states $D^{(2)}(\omega)$ of the 36-site VBS$_3$ state.  Each column shows $\omega_{\tau,b}({\bf k})$, $\sigma^{(2)}({\omega})$, and $D^{(2)}(\omega)$ for a given value of $D/J$.  In the plots of the triplon dispersions, the $\tau=+,0,-$ dispersions are drawn with blue, green, magenta lines, respectively.  When $D/J=0$, the blue, green, and magenta lines are overlapped due to the fact that the $\tau=+,0,-$ dispersions are all degenerate.  In the plots of the optical conductivity and density of states, we only plot the $+-$ component (blue) and total sum (red) of the $+-$ and 00 components to avoid confusion. \label{fig:36_VBS_3_band_conductivity}}
\end{figure*}

In the case of the 36-site VBS$_3$ state in Fig. \ref{fig:VBS_patterns} (c), overall patterns in the triplon dispersions, optical conductivity, and density of states are quite similar to the previous cases with several important differences. In this case, even for $D=0$, there is substantial triplon dispersion as shown in Fig. \ref{fig:36_VBS_3_band_conductivity} (a). Thus, even for $D=0$, the two triplon density of states is not composed of pure delta-functions and this is directly reflected in the optical conductivity. As $D/J$ is increased, similar to other cases, $D^{(2)}(\omega)$, and consequently $\sigma^{(2)}(\omega)$ lose sharp peak-like structures. On the other hand, in the 36-site VBS$_3$ state triplon dispersion is less degenerate than in the previous two VBS phases as revealed by DM interaction. When the DM interaction is absent, $\omega_{\tau,b}({\bf k}) ~ (\tau=\pm,0)$ are all degenerate due to the SO(3) spin-rotation symmetry (see \ref{fig:36_VBS_3_band_conductivity} (a)). If the DM interaction is, however, turned on, the degeneracy is lifted as $\omega_{+,b}({\bf k})\ne\omega_{-,b}({\bf k})$ due to the lack of inversion symmetry (see magneta and blue lines in Fig. \ref{fig:36_VBS_3_band_conductivity} (b)$\sim$(d)) unlike in the other two VBS phases where $\omega_{+,b}({\bf k})=\omega_{-,b}({\bf k})$ holds because of the inversion and time reversal symmetries.

While some of the details of the peak shapes may actually change when fluctuations beyond mean-field approximation are taken into account, we notice the startling qualitative feature that in all the three cases considered, the sharp structures in optical conductivity is increasingly replaced with many closed spaced peaks. This is a reflection that the triplon band structure becomes more dispersive with increasing DM interaction. Interestingly, similar effects have recently been observed even in certain spin liquid states on the Kagome lattice where the effect of the DM term is to increase the spinon dispersion \cite{2013_Dodds}. In those cases, the resulting dynamic spin-structure factor becomes more and more diffused.

In the present case, in absence of sufficient experimental resolution these peaks may appear a broad feature in optical conductivity, which increases as the DM term is increased. The resulting envelope of the peak structure may even mimic a power-law response within a limited range of frequency. Such power-law responses are expected in certain U(1) spin-liquid states \cite{2007_Ng,2013_Potter,2013_Pilon}.

Another situation that needs to be pointed out is: within bond operator mean-field theory, a second order  transition from VBS to a magnetically ordered state occurs when the single-triplon gap closes continuously leading to triplon condensation \cite{2004_Matsumoto,2008_cepas,2012_Hwang}. Close to such a transition, on the VBS side, the single-triplon gap is small and hence we expect the broad hump, as seen in all the three VBSs above, to shift to lower energy and most likely ultimately replaced by power-law behaviour right at the critical point. This situation should be contrasted with the other more exotic phase transitions that have been recently discussed \cite{2013_Potter,2013_Huh}.

While it is not clear, to the best of our knowledge, what kind of spin Hamiltonian on the Kagome lattice may yield such a transition, we note that recent  variational Monte Carlo simulation studies \cite{2011_Iqbal} show addition of small but finite ferromagnetic next nearest neighbor (NNN) coupling to the nearest neighbour Heisenberg antiferromagnet on the ideal Kagome lattice stabilizes a VBS state. Similarly, another calculation \cite{2008_cepas} suggest that a N\'eel phase is obtained by tuning the DM interactions of the form considered here. We can speculate that in the two-dimensional phase diagram containing both ferromagnetic NNN and DM, such a transition may be possible. 

\section{Summary and outlook\label{sec:last_sec}}

We now summarize our results. In this work, we have explored the subgap optical conductivity in Mott insulators in presence of DM interactions by extending the formalism in Ref. \cite{2008_Bulaevskii}. In Mott insulators, such subgap conductivity results from finite electronic polarization generated by virtual charge fluctuations. The polarization then couple to a uniform external electric field to give finite optical response much below the single-particle charge gap. We have then applied the above formalism to investigate the optical conductivity of various VBS states on both ideal and deformed Kagome lattices. We have used the well-known bond operator mean field theory to characterize the low energy triplon excitations of the VBS phase and the leading order contribution to optical conductivity results from two-triplon-excited states.

For the deformed Kagome lattice, we work out the results for the specific compound Rb$_2$Cu$_3$SnF$_{12}$ that has a {\it pinwheel} VBS ground state with 12-site unit cell. Using a quantitatively accurate low energy triplon spectra, we obtain the optical conductivity to figure out the different energy scale and the peak structures. The minimum frequency of the optical response is bounded by the minima of the two-triplon excitation gap. Future experiments on the optical conductivity of this material will provide useful insights regarding the accuracy of our calculation.

On the ideal Kagome lattice, we investigate the optical conductivity of three different VBS solids: one with 12-site unit cell and two with 36-site unit cell for various values of the DM interactions. As expected, we find the two-triplon excitation gap provides the lower cut-off for the optical response. However, as one increases the DM interactions, the sharp features in the optical conductivity are replaced by a smooth and broad envelope resembling incoherent response. It is important to distinguish such broad hump like behaviour in experiments from the power-law optical response expected in  U(1) spin liquids, particularly when the two-triplon gap is small. Such small triplon gap may arise in a VBS near transition with magnetically ordered state.

Lastly, we note that recently the magneto-elastic mechanism of coupling to an external electric field has also been discussed \cite{2008_Bulaevskii,2013_Potter,2013_Huh}. In this mechanism, the electric field distorts the lattice to modulate the spin exchange couplings which then lead to finite electric polarization and hence optical response. In this work, we have assumed a rigid lattice and hence neglected the contributions to optical conductivity coming from the magneto-elastic coupling. While the form of the polarization operators are not expected to change (because they are constrained by symmetry), the prefactors are different and are controlled by different elastic moduli of the system \cite{2008_Bulaevskii}. This forms an interesting avenue for future research, particularly the comparison of these two contributions for a given compound can be interesting in regards to experiments. Another point that deserves mention in this regard is the pure phonon contribution to the optical conductivity, usually occuring above 3 THz frequency scale, needs to be carefully separated from the present spin-contribution \cite{2011_Basov}.


\section{Acknowledgements}

The authors acknowledge SungBin Lee and Tyler Dodds for useful discussions. This research was supported by the NSERC, CIFAR, and Centre for Quantum Materials at the University of Toronto.

\appendix

\section{Fitting\label{appendix:fitting}}

For a quantitatively correct prediction for the optical conductivity of the compound Rb$_2$Cu$_3$SnF$_{12}$, 
we fit the triplon spectrum from the quadratic Hamiltonian in Eq. (\ref{eq:H_quad}) into the experimental results
by considering renormalizations of the parameters $J_n, ~ D_n ~ (n=1,\cdots,4)$.
The fitting is done in such a way that the four characteristic excitation energies $\Delta_n ~ (n=0,\cdots,3)$ in Fig. \ref{fig:triplon_dispersion} (a)
have following correct energy scales \cite{2010_Matan}:
\begin{equation}
 \begin{array}{lrlrlrlr}
 \Delta_0^{NS} & = & 2.35 & \textup{meV} , & \Delta_1^{NS} & = & 7.3 & \textup{meV} ,
 \\
 \Delta_2^{SE} & = & 5.1 & \textup{meV} , & \Delta_3^{SE} & \simeq & 4.5 & \textup{meV} ,
 \end{array}
 \label{eq:four_spectra}
\end{equation}
where the superscripts NS and SE mean the neutron scattering and series expansion, respectively.
The resultant triplon excitation spectra is shown in Fig. \ref{fig:triplon_dispersion} (b).
\section{Table for ${\bf M}$}
\label{appen_table}
\begin{table}
\centering
\begin{tabular}{c|ccc|ccc|c|c|c}
\hline
link & $m$ & $a$ & ${\bf R}$ & $l$ & $b$ & ${\bf R}'$ & $M_{ij}^x$ & $M_{ij}^y$ & $d_{ij,k}$
\\
\hline
$J_1$& 1 & $L$ & ${\bf 0}$ & 1 & $R$ & ${\bf 0}$ & $-A$ & $-A/\sqrt{3}$ & $d_1$
\\
& 2 & $L$ & ${\bf 0}$ & 2 & $R$ & ${\bf 0}$ & 0 & $-2A/\sqrt{3}$ & $d_1$
\\
& 3 & $L$ & ${\bf 0}$ & 3 & $R$ & ${\bf 0}$ & $A$ & $-A/\sqrt{3}$ & $d_1$
\\
& 4 & $L$ & ${\bf 0}$ & 4 & $R$ & ${\bf 0}$ & $A$ & $A/\sqrt{3}$ & $d_1$
\\
& 5 & $L$ & ${\bf 0}$ & 5 & $R$ & ${\bf 0}$ & 0 & $2A/\sqrt{3}$ & $d_1$
\\
& 6 & $L$ & ${\bf 0}$ & 6 & $R$ & ${\bf 0}$ & $-A$ & $A/\sqrt{3}$ & $d_1$
\\
\hline
$J_2$& 1 & $R$ & ${\bf 0}$ & 2 & $R$ & ${\bf 0}$ & 0 & $2A/\sqrt{3}$ & $d_2$
\\
& 2 & $R$ & ${\bf 0}$ & 3 & $R$ & ${\bf 0}$ & $-A$ & $A/\sqrt{3}$ & $d_2$
\\
& 3 & $R$ & ${\bf 0}$ & 4 & $R$ & ${\bf 0}$ & $-A$ & $-A/\sqrt{3}$ & $d_2$
\\
& 4 & $R$ & ${\bf 0}$ & 5 & $R$ & ${\bf 0}$ & 0 & $-2A/\sqrt{3}$ & $d_2$
\\
& 5 & $R$ & ${\bf 0}$ & 6 & $R$ & ${\bf 0}$ & $A$ & $-A/\sqrt{3}$ & $d_2$
\\
& 6 & $R$ & ${\bf 0}$ & 1 & $R$ & ${\bf 0}$ & $A$ & $A/\sqrt{3}$ & $d_2$
\\
\hline
$J_4$& 1 & $R$ & ${\bf 0}$ & 6 & $L$ & ${\bf 0}$ & 0 & $-2A/\sqrt{3}$ & $d_4$
\\
& 2 & $R$ & ${\bf 0}$ & 1 & $L$ & ${\bf 0}$ & $A$ & $-A/\sqrt{3}$ & $d_4$
\\
& 3 & $R$ & ${\bf 0}$ & 2 & $L$ & ${\bf 0}$ & $A$ & $A/\sqrt{3}$ & $d_4$
\\
& 4 & $R$ & ${\bf 0}$ & 3 & $L$ & ${\bf 0}$ & 0 & $2A/\sqrt{3}$ & $d_4$
\\
& 5 & $R$ & ${\bf 0}$ & 4 & $L$ & ${\bf 0}$ & $-A$ & $A/\sqrt{3}$ & $d_4$
\\
& 6 & $R$ & ${\bf 0}$ & 5 & $L$ & ${\bf 0}$ & $-A$ & $-A/\sqrt{3}$ & $d_4$
\\
\hline
$J_3$& 1 & $L$ & ${\bf 0}$ & 5 & $L$ & ${\bf r}_C$ & $B$ & $B/\sqrt{3}$ & $d_3$
\\
& 2 & $L$ & ${\bf 0}$ & 6 & $L$ & $-{\bf r}_A$ & $0$ & $2B/\sqrt{3}$ & $d_3$ 
\\
& 3 & $L$ & ${\bf 0}$ & 1 & $L$ & $-{\bf r}_B$ & $-B$ & $B/\sqrt{3}$ & $d_3$
\\
& 4 & $L$ & ${\bf 0}$ & 2 & $L$ & $-{\bf r}_C$ & $-B$ & $-B/\sqrt{3}$ & $d_3$
\\
& 5 & $L$ & ${\bf 0}$ & 3 & $L$ & ${\bf r}_A$ & 0 & $-2B/\sqrt{3}$ & $d_3$
\\
& 6 & $L$ & ${\bf 0}$ & 4 & $L$ & ${\bf r}_B$ & $B$ & $-B/\sqrt{3}$ & $d_3$
\\
\hline
\end{tabular}
\caption{The coefficients $\{ {\bf M}_{ij} (=M_{ij}^x\hat{x}+M_{ij}^y\hat{y}) \}$ and $\{ d_{ij,k} \}$ for the 12-site unit cell of the deformed Kagome lattice antiferrmagnet. The sites $i,j$ are denoted with $i=(m,a,{\bf R})$ and $j=(l,b,{\bf R}')$, where $m,l(=1,\cdots,6)$ are dimer indices, $a,b=L,R$, and ${\bf R},{\bf R}'$ are lattice vectors.
In our convention, ${\bf S}_R$ in a dimer is located at the vertices of the $J_2$-link hexagon in Fig. \ref{fig:deformed_kagome_lattice}.
\label{tab:L}}
\end{table}

In the table, the parameters $\{d_{ij,k}\}$ in Eq. (\ref{eq:I_ij}) for the bonds in the unit cell are given in terms of $d_1,d_2,d_3,d_4$, where
\begin{equation}
 \begin{array}{llllllll}
 d_1 & = & -\frac{D_1}{J_1}+\frac{D_2}{J_2}+\frac{D_4}{J_4}, & d_2 & = & \frac{D_1}{J_1}-\frac{D_2}{J_2}+\frac{D_4}{J_4},
 \\
 d_4 & = & \frac{D_1}{J_1}+\frac{D_2}{J_2}-\frac{D_4}{J_4},  & d_3 & = & \frac{D_3}{J_3}.
 \end{array}
\end{equation}


\section*{References}

\end{document}